\documentclass[12pt,a4paper]{article}
\usepackage{amsmath,amssymb}
\usepackage{graphicx}
\usepackage{array}
\usepackage{wasysym}

\makeatletter
\newif\if@preliminary
\@preliminaryfalse
\def\preliminary{\@preliminaryfalse}

\def\bq{\begin{equation}}
\def\eq{\end{equation}}
\def\ba{\begin{eqnarray}}
\def\ea{\end{eqnarray}}

\def\preprintno#1{\def\@preprintno{#1}}
\def\address#1{\def\@address{#1}}
\def\email#1#2{\thanks{\tt #1@{}#2}}
\def\abstract#1{\def\@abstract{#1}}
\renewcommand\abstractname{ABSTRACT}
\newlength\preprintnoskip
\newlength\abstractwidth
\@titlepagetrue
\renewcommand\maketitle{\begin{titlepage}
  \let\footnotesize\small
  \hfill\parbox{\preprintnoskip}{
  \begin{flushright}\@preprintno\end{flushright}}\hspace*{1cm}
  \vskip 60\p@
  \begin{center}
    {\Large\bf\boldmath \@title \par}\vskip 1cm
    {\sc\@author \par}\vskip 3mm
    {\@address \par}
    \if@preliminary
      \vskip 2cm {\large\sf PRELIMINARY DRAFT \par \@date}
    \fi
  \end{center}\par
  \@thanks
  \vfill
  \begin{center}
    \parbox{\abstractwidth}{\centerline{\abstractname}
    \vskip 3mm
    \@abstract}
  \end{center}
  \end{titlepage}
  \setcounter{footnote}{0}
  \let\thanks\relax\let\maketitle\relax
  \gdef\@thanks{}\gdef\@author{}\gdef\@address{}
  \gdef\@title{}\gdef\@abstract{}\gdef\@preprintno{}
}

\def\@citex[#1]#2{\if@filesw\immediate\write\@auxout{\string\citation{#2}}\fi
  \def\@citea{}\@cite{\@for\@citeb:=#2\do
    {\@citea\def\@citea{,\penalty\@m}\@ifundefined
       {b@\@citeb}{{\bf ?}\@warning
       {Citation `\@citeb' on page \thepage \space undefined}}%
\hbox{\csname b@\@citeb\endcsname}}}{#1}}
\def\citerange{\@ifnextchar [{\@tempswatrue\@citexr}{\@tempswafalse\@citexr[]}}
\def\@citexr[#1]#2{\if@filesw\immediate\write\@auxout{\string\citation{#2}}\fi
  \def\@citea{}\@cite{\@for\@citeb:=#2\do
    {\@citea\def\@citea{--\penalty\@m}\@ifundefined
       {b@\@citeb}{{\bf ?}\@warning
       {Citation `\@citeb' on page \thepage \space undefined}}%
\hbox{\csname b@\@citeb\endcsname}}}{#1}}
\long\def\@makecaption#1#2{
  \vskip\abovecaptionskip
  \sbox\@tempboxa{#1: \emph{#2}}
  \ifdim \wd\@tempboxa >\hsize
    #1: \emph{#2}\par
  \else
    \hbox to\hsize{\hfil\box\@tempboxa\hfil}
  \fi
  \vskip\belowcaptionskip}

\def\fmslash{\@ifnextchar[{\fmsl@sh}{\fmsl@sh[0mu]}}
\def\fmsl@sh[#1]#2{
  \mathchoice
    {\@fmsl@sh\displaystyle{#1}{#2}}
    {\@fmsl@sh\textstyle{#1}{#2}}
    {\@fmsl@sh\scriptstyle{#1}{#2}}
    {\@fmsl@sh\scriptscriptstyle{#1}{#2}}}
\def\@fmsl@sh#1#2#3{\m@th\ooalign{$\hfil#1\mkern#2/\hfil$\crcr$#1#3$}}
\makeatother

\newcommand{\GeV}{{\ensuremath\rm GeV}}

\newcommand{\cond}{\ensuremath{\stackrel{!}{=}}}

\newcommand{\textindex}[1]{{\mbox{\scriptsize #1}}}

\newcommand{\feynarts}{\texttt{FeynArts}}
\newcommand{\formcalc}{\texttt{FormCalc}}
\newcommand{\vamp}{\texttt{VAMP}}

\begin{document}

\preliminary        


\preprintno{IPPP/09/97, DCTP/09/194\\Edinburgh 2009/17\\SI-HEP-2009-16\\[0.5\baselineskip] 
\today}

\title{Numerical Evaluation of Feynman Loop Integrals by Reduction to Tree Graphs
}

\author{ W.~Kilian\email{kilian}{physik.uni-siegen.de}$^a$,
 T.~Kleinschmidt\email{tobias.kleinschmidt}{durham.ac.uk}$^{bc}$}

\address{\it
$^a$ University of Siegen, Fachbereich Physik, D--57068 Siegen, Germany \\
$^b$ University of Edinburgh, Institute for Particle and Nuclear Physics, EH9 3JZ, UK\\
$^c$ University of Durham, Institute for Particle Physics Phenomenology, DH1 3LE, UK}

\abstract{We present a new method for the numerical evaluation of loop
  integrals which is based on the Feynman Tree Theorem.  The
  loop integrals are replaced by phase-space integration over
  fictitious extra on-shell particles.  This integration can be
  performed alongside with the Monte-Carlo integration of ordinary
  phase space, avoiding the time-consuming nesting of loop evaluation
  inside the integrand, and directly leading to NLO event generation.
  We systematically construct subtractions, necessary to cancel both
  ultraviolet divergences and the extra threshold singularities in
  phase-space which arise in the numerical evaluation.  Infrared
  singularities can be dealt with by standard methods.
  As a proof
  of concept, we apply the method to NLO Bhabha scattering in QED
  and construct the corresponding NLO Monte Carlo
  event generator.  }

\maketitle


\section{Introduction}

The advent of the LHC has initiated a strong activity in the
development of new methods and tools aiming at an efficient
computation of processes at next-to-leading order (NLO) in
perturbation theory.  Automated programs exists which simulate events
at leading order (LO) with many (typically, up to six or eight)
final-state partons without factorizing them into production and
decay \citerange{Mangano:2002ea,Cafarella:2007pc}. 
These can be interfaced and properly matched to parton-shower
algorithms and thus provide a powerful tool for the simulation of
events with complex final states at particle colliders.  For specific
applications, this program has already been extended to NLO processes.

Especially at hadron colliders, the LO predictions are highly
scale-dependent and lack important contributions, so it becomes
increasingly important to simulate proper events at the NLO level.

The cancellation of IR divergences and the inclusion of subtraction
terms in NLO calculations are well understood in principle, and their
implementation in universal event generators is in progress.  A
crucial problem of NLO event generation originates from the loop
contributions to multi-parton matrix elements. The number of
individual Feynman graphs rises dramatically with the number of
external legs, and tensor reduction methods increase the number of
terms even more.  This poor scaling behavior, and the associated
instabilities due to large numerical cancellations in matrix elements,
make it worthwhile to investigate different and alternative approaches
to the problem of automatic NLO computation and simulation.

A particular alternative approach to evaluating amplitudes at the loop
level is to exploit their analytic properties. Observing that any
amplitude can be decomposed into cut contributions and a simple
rational part, one can construct the coefficients of a set of basic
integrals by cutting lines of the loop graph in a particular way and
evaluating the resulting tree-level amplitudes.  In the recent past,
some $2\rightarrow 4$ processes for the LHC have been computed using
either of these two main techniques \citerange{Bredenstein:2009aj,Reiter:2009kb},
and there are several
groups aiming at a full automatization of NLO processes.

In this paper, we propose a new method for the evaluation of loop
integrals, which allows for direct numerical computation without
reduction to a set of basic integrals.  The matrix element is
re-expressed using an improved version of the Feynman Tree Theorem
(FTT) \citerange{Feynman:1963ax,Feynman:1972mu}.
Roughly speaking, this theorem relates loop graphs to tree
graphs, and it is nothing but the relation of relativistic
Feynman-graph perturbation theory to ``old-fashioned''
non-relativistic perturbation theory.  It holds in any local
relativistic field theory. The loop integrals are transformed into
phase space integrals and are evaluated numerically along with the
phase space integral over of the external partons of the process under
consideration.  So, there is only one step of numerical integration
involved in the computation of any particular integrated cross section
or distribution.  We therefore can make use of powerful existing
technologies for numerical phase-space integration by tree-level event
generators to evaluate processes at the one-loop level.

There has recently been increased interest in the Feynman tree
theorem.  In \cite{Brandhuber:2005kd} it was used to prove the
covariance of non-MHV amplitudes at one-loop level obtained from MHV
diagrams, as well as for the calculation of MHV scattering amplitudes
in $\mathcal{N}=4$ super Yang-Mills theory. In \cite{NigelGlover:2008ur}, 
single cuts were used to simplify one-loop scattering amplitudes
in massless gauge theories.

An algorithm for the direct numerical integration of NLO processes by
the FTT method was developed first for massless amplitudes by Soper
\textit{et al.}~\cite{Soper:1998ye,Nagy:2006xy}.  There, internal
on-shell or threshold singularities were found to be a serious
obstacle to numerical phase-space integration.  To avoid these
singularities, contour deformation of the integration into the complex
plane was used, either in three-momentum space or in Feynman-parameter
space.

In the present work we allow for massive particles.  Threshold
singularities are handled by a specially taylored subtraction
procedure that allows us to stay in the space of real integration
parameters.  This enables us to directly implement the integrands in
an ordinary Monte Carlo event generator, and it also gives a handle on
the treatment of overlapping singularities in the integration region,
which can occur in integrals from six external legs on.

In a project independent of the present one, Catani \textit{et
  al.}~\cite{Catani:2008xa} exploited the relation between loop
integrals and phase space integrals to express loop amplitudes as sum
of tree amplitudes arising solely from single cuts.  Contributions
from multiple cuts which are present in the Feynman Tree Theorem, are
compensated by a non-trivial $i\epsilon$-prescription in the
propagators.

The paper is structured as follows. In a first technical part, we
develop the method. In section \ref{s_ftt}, we review the derivation
of the Feynman Tree Theorem and provide an improved version where the
initial $i\epsilon$ prescription disappears, allowing for a direct
numerical integration. In the following sections \ref{s_ren_reg} to
\ref{s_uv/ir}, we propose renormalization and regularization schemes
and discuss the treatment of infrared divergences. We then discuss the
types of internal singularities that can arise in the numerical
evaluation and present the construction of appropriate subtraction
function in Sec.~\ref{s_threshold_singularities}. In the next part,
section \ref{s_proof_of_concept}, we discuss the implementation of
this method in a Monte Carlo cross section integration and event
generation for Bhabha scattering in QED. We thereafter conclude and
give an outlook.\footnote{Preliminary studies for the present project
  can be found in~\cite{Diplomarbeiten}.  Additional details of the
  techniques involved are presented in~\cite{PhD thesis}.}


\section{The Method}\label{s_method}

\subsection{Feynman Tree Theorem}\label{s_ftt}

The starting point of our approach is a theorem by R.~Feynman
\citerange{Feynman:1963ax,Feynman:1972mu}, which was
formally stated in order to invest the renormalizability of a quantum
theory of gravitation.  The idea is to decompose loop propagators into
advanced Green functions and a delta function. When integrating over
the zero component of the loop momentum, the delta functions will set
the internal momentum of the associated propagator on-shell. This has
the effect of {\it opening} or {\it cutting} the loop. The Feynman
Tree Theorem states, that a loop integral can be expressed as the sum
of all possible cuts of its propagators. Terms with one cut propagator
can be interpreted as tree-level processes with an additional outgoing
and an additional \emph{incoming} particle with identical
momentum. The original integration over the loop momentum becomes a
phase space integration for the additional particle.  This extra
integration is peculiar since its phase space is not restricted by the
available process energy. 

Operating on every loop diagram, a full one-loop $m\rightarrow n$
matrix element can be rewritten as the coherent sum of all possible $m
\rightarrow n+p+\bar{p}$ tree level processes with an additional
integration over the phase space of the particle $p$ and its
corresponding antiparticle $\bar{p}$ obtained by crossing from the
initial state.

\subsubsection{Derivation}
In the following, we will briefly review the derivation of the Feynman
Tree Theorem, closely following \cite{Feynman:1972mt} (see
also~\cite{Brandhuber:2005kd}). In the next section, we give a
modified version of the Tree Theorem which is better suited for
numerical integration.

The integrand $I(k)$ of a loop integral can be written as a product of
Feynman Green functions $F$ of the Klein-Gordon operator times a
regular function $N(k)$ in the numerator. The latter may depend on the
integration momentum $k$ and have Lorentz and Dirac indices but is not
of interest in the following. Suppressing the possible indices, we
have
\begin{equation}\label{integrand}
I(k)=N(k)\prod_{i}F(k+p_i,m_i),
\end{equation} 
\noindent where we used the letter $\mbox{F}$ to indicate the use of
the Feynman prescription. The $p_i$ are linear combinations of
external momenta, the $m_i$ the masses of the physical particle the
propagator corresponds to. We define
\begin{equation}\label{fprop}
F_i\equiv F(k+p_i,m_i)
=\frac{i}{(k+p_i)^2-m_i^2+i\epsilon}.
\end{equation}
Note that we only consider cases where the $k$-dependence in the
denominator of a propagator is of the form (\ref{fprop}). Here and in
the following, we use 't Hooft-Feynman gauge, where gauge boson
propagators take on this simple structure in the denominator.
                             
In the following, we want to replace the Feynman Green functions $F_i$
by advanced Green functions $A_i$, where both poles lie in the upper
$k^0$-half complex plane. We define
\begin{equation}
E_i=\sqrt{(\vec{k}+\vec{p}_i)^2+m_i^2}.
\end{equation}
Performing a partial-fraction decomposition in (\ref{fprop}), and
similar for $A_i$, we get
\begin{eqnarray}
F_i&=&\frac{i}{2(E_i-i\epsilon)}\left(\frac{1}{k^0-(-p_i^0+E_i)+i\epsilon}-\frac{1}{k^0-(-p_i^0-E_i)-i\epsilon}\right),\label{F}\\
A_i&=&\frac{i}{2E_i}\left(\frac{1}{k^0-(-p_i^0+E_i)-i\epsilon}-\frac{1}{k^0-(-p_i^0-E_i)-i\epsilon}\right).\label{A}
\end{eqnarray}
Using a representation of the delta function
\begin{equation}\label{delta_rep}
2\pi i\delta(u)=\lim\limits_{\epsilon\rightarrow +0}\left(\frac{1}{u-i\epsilon}-\frac{1}{u+i\epsilon}\right),
\end{equation}
\noindent the difference of the Feynman and advanced Green function
$F_i-A_i$ is given by:
\begin{equation}\label{Deltal}
\Delta_i^l=\frac{2\pi}{2E_i}\delta(k^0-(-p^0_i+E_i)).
\end{equation}
Here, we ignored the imaginary part in the prefactor of the Feynman
Green function (\ref{F}). Being independent of $k^0$, it is not
relevant in the following. The superscript $l$ indicates that this
delta function picks out the propagator pole, which originally was
situated in the lower half plane, setting the momentum $k+p_i$ on its
mass shell with a positive energy component. The $k^0$ integration
over a product of advanced Green functions $A_i$ vanishes, since for
two or more Green functions the integrand falls of sufficiently fast
for large $k^0$.  Thus, one can close the contour of integration in
the lower half plane where no poles are situated:
\begin{equation}
0\!=\!\! \int\! N(k)\prod_{i}^n A_i.
\end{equation}
Replacing $A_i$ with $F_i-\Delta^l_i$, we get:
\begin{equation}
0=\!\! \int \! N(k)\left[F \cdots F - \sum \Delta^l F \cdots +\sum \Delta^l \Delta^l F\cdots 
-\ldots +(-1)^{n} \sum \Delta^l\cdots\Delta^l\right],\label{ftt_original}
\end{equation}
\noindent where we skipped indices in the terms in the brackets.

Equation (\ref{ftt_original}) is the Feynman Tree Theorem
\cite{Feynman:1963ax,Feynman:1972mt}.  Since the delta functions
cancel the integration, it states that a loop integral can be
expressed as a sum of tree amplitudes. The first sum runs over all
permutations where one propagator is replaced by a delta function, the
second sum runs over all terms including two delta functions, and so
on. The first term on the right-hand side, which contains only Feynman
Green functions, is the original integrand (\ref{integrand}). In the
following the term {\it cutting a propagator} of a loop will refer to
one of the terms in (\ref{ftt_original}), where a propagator was
replaced by a delta function.

In \cite{Catani:2008xa}, equation (\ref{ftt_original}) was the
starting point for constructing a method that re-expresses loop
amplitudes as a sum of tree amplitudes that involve single cuts only.
The cost of this simplification lies in an exchange of the Feynman
propagators (\ref{F}) by propagators that acquire a rather complicated
$i\epsilon$ prescription.

To make use of the FTT in a numerical evaluation of loop integrals
involving only real numbers, at some point we have to set the
$i\epsilon$ terms in the denominators to zero.  Clearly, this can not
be done in a na\"ive way, since terms including the delta functions
$\Delta^l$ were derived precisely from a configuration of Feynman and
advanced Green functions with a unique description of the poles in the
complex $k^0$-plane.

\subsubsection{Improved Version}
We can re-express the Feynman Green functions $F_i$ by use of the
identity:
\begin{equation}\label{principal_value}
\frac{1}{x-a\pm i\epsilon}=\mathcal{P}\frac{1}{x-a}\mp i\pi\delta(x-a),
\end{equation}
\noindent where $\mathcal{P}$ is Cauchy's Principal Value, which is
obtained by evenly approaching the singular point from both sides such
that the diverging pieces cancel each other. Applying this identity to
the Feynman Green function (\ref{F}) and again skipping the
$i\epsilon$ term in the factor in front of the brackets we get:
\begin{eqnarray}
F_i&=&\frac{i}{2E_i}\left(\mathcal{P}\frac{1}{k^0-(-p_i^0+E_i)}-i\pi\delta(k^0-(-p_i^0+E_i))\right.\nonumber\\
&&\hspace{1.5cm}\left. -\mathcal{P}\frac{1}{k^0-(-p_i^0-E_i)} -i\pi\delta(k^0-(-p_i^0-E_i))\right)\nonumber\\
&\equiv &P_i+\frac{1}{2}\Delta^l_i+\frac{1}{2}\Delta^u_i ,\label{P}
\end{eqnarray}
\noindent where we defined $\Delta^u_i$ as the delta function that
sets the zero component of the negative momentum of the associated
propagator $i$:
\begin{equation}
\Delta^u_i = \frac{2\pi}{2E_i}\delta(k^0-(-p^0_i-E_i)).
\end{equation}
\noindent $P_i$ stands for the propagator with no
$i\epsilon$-prescription in the denominator:
\begin{equation}
P_i=\mathcal{P}\frac{i}{(k+p_i)^2-m_i^2}.
\end{equation}
In numerical evaluations of tree amplitudes, propagators of this form,
without $i\epsilon$-terms, are used. Inserting (\ref{P}) in
(\ref{ftt_original}), we therefore get after some combinatorics a
version of the Tree Theorem (\ref{ftt_original}) which is better
suited for numerical evaluations:
\begin{eqnarray}
\int\!I(k)\!\!\!&=\!\!\!&\int N(k)\big[
\Delta^l_1 P_2\cdots P_n + P_1\Delta^l_2 P_3\cdots P_n + \ldots + 
P_1\cdots P_{n-1}\Delta^l_n\big]\nonumber\\
&&\vspace{1mm}\nonumber\\
&&+\int\!N(k)\!\!\!\!\!\sum\limits_{\mbox{\tiny $\begin{array}[t]{c}perm.\\U+L\ge 2\end{array}$}}\!\!\!\!\!\! C_{LUP}\,\,\Delta^{l^L}\Delta^{u^U} P^P, \label{ftt}\\
&&\nonumber\\
\! C_{LUP}\!\!\!&=\!\!\!&\frac{1}{2^{L+U}}\left(1-(-1)^L\right)\label{coeff}.
\end{eqnarray}
Since the structure of (\ref{ftt}) is still very similar to
(\ref{ftt_original}), we will from now on refer to (\ref{ftt}) as the
Feynman Tree Theorem. 

The sum runs over all possible permutations, where the functions
($\Delta^l$,$\Delta^u$,$P$) appear ($L$,$U\!$,$P$) times, with the
additional constraint $L+U+P=n$. The coefficient $C_{LUP}$ stands in
front of every term. Note that terms with an even number of $\Delta^l$
functions vanish. This is a generalization of the observation that a
loop integral does not get an imaginary part at a momentum
constellation where two poles in the lower $k^0$-half plane
coincide. This is in contrast to a pinch singularity, where the
contour of integration is trapped between poles in the lower and upper
half plane. We will discuss these contributions in more detail in
section \ref{s_threshold_singularities}.
 
We explicitly wrote out the terms containing one $\Delta^l$ function
in the first line of (\ref{ftt}). Here, in each term, one of the
propagators of the original loop is replaced by a delta
function. After $k^0$ integration, all of these terms can be
interpreted as tree graphs with one additional incoming and outgoing
particle and an additional phase space integral over this particles
momentum,
\begin{equation}
\int\!\! \frac{d^3k}{(2\pi)^3 2E_i}.
\end{equation}
Note however, that the momentum which is put on-shell is $k+p_i$ and
the integration is performed over $k$. In general, one must not shift
the integration momentum in a single term only, since the integrand
consists of several terms which are coherently summed up. Some of
these terms may have peaks or may even be UV divergent, since we will
not use dimensional regularization. Only in the sum of the individual
tree graphs, these singularities will then be cancelled.

When a propagator is replaced by $\Delta^l_i$, setting $k+p_i$ on the
mass shell, the remaining numerator can be read as the product of one
additional incoming and outgoing external on-shell particle,
\begin{eqnarray}
(\fmslash k +\fmslash p_i+m)=\sum\limits_\lambda u_\lambda(k+p_i)\bar{u}_\lambda(k+p_i);\\
(\fmslash k +\fmslash p_i-m)=\sum\limits_\lambda v_\lambda(k+p_i)\bar{v}_\lambda(k+p_i);\\
-g_{\mu\nu}=\sum\limits_\sigma \epsilon^*_\mu(k+p_i;\sigma)\epsilon_\nu(k+p_i;\sigma),
\end{eqnarray}
\noindent where the sum runs over all physical and unphysical internal
states. For a cut fermion propagator, we will obtain the particle if
the momentum flow of the loop is in the same direction as the fermion
number flow, and the antiparticle otherwise.

\subsubsection{Construction of graphs}

Thus, any loop graph can be re-expressed as a sum of tree graphs.
Turning this around, one can create the complete set of tree graphs
with an inclusive incoming and outgoing particle, and sew these loose
ends together to obtain loop diagrams again.  One thus finds all
one-loop corrections to a given $2\rightarrow n$ process by
considering all possible tree graphs with two additional particles,
writing schematically
\begin{equation}\label{looptotree}
\mathcal{M}^{\textindex{1-loop}}(2\rightarrow n) \Longleftrightarrow \sum\limits_{X}\mathcal{M}^{\textindex{Tree}}(2\rightarrow n + X + \bar{X}) + \ldots .
\end{equation}
Here, the sum runs over the particle content of the theory, and we
flipped the initial state particle to the corresponding antiparticle
in the final state, using crossing symmetry. 

In principle, this expression includes also tadpole diagrams and
self-energy corrections to external on-shell legs.  These may be set
to zero by appropriate renormalization conditions; in that case, they
can be ignored in the right-hand side of (\ref{looptotree}).

The creation of all relevant Feynman graphs contributing to a process
in a given order is therefore reduced to the task of creating
tree-level graphs with the corresponding additional particles.  For
this task, powerful tools exist. However, we cannot always shift
momentum in the individual tree graphs freely, since as stated above,
some graphs may have ultraviolet divergences which only in the
coherent sum of the tree graphs cancel. Therefore, we cannot fully
exploit the reduction mechanisms implemented in these tools.

\subsubsection{Monte-Carlo integration}

The remaining three-dimensional integral over the loop momentum can
now be pulled out of the individual graphs and put in front of the
amplitude, together with the phase space integrals over the external
particles which is present in a cross-section calculation.  Since the
extra integration is also of the form of a phase-space integral,
techniques developed for the integration, in particular multi-channel
sampling, can immediately be adopted.  Furthermore, this integration
can be performed \emph{simultaneously} with the external phase space
integration. 

Using a suitable Monte Carlo integration routine, going from
Born-level processes to loop level merely amounts to an increase of
the integration dimension.  If integration mappings and multi-channel
weights can be successfully adapted to this situation, this results
just in a minor increase in the number of sampling points, if one
wants to achieve a NLO precision at the same level as the LO
calculation.\footnote{Of course, the calculation now involves
  amplitudes with an increased number of external legs, so it is
  essential to use a matrix-element generator with optimal scaling
  behavior.}

From the viewpoint of Monte-Carlo event generation, each sampling
point corresponds to a particular configuration of external \emph{and}
internal momenta in the original loop amplitude, and the accumulation
of sampling points resolves both loop and phase-space integration at
the same time.  This is in contrast to traditional analytical methods,
where for each individual configuration of external momenta the
analytical result of the loop integration has to be numerically
evaluated.  This task normally includes a time-costly calculation of a
large number of polylogarithms, and it is subject to numerical
instabilities associated to Gram determinants, as an artifact of
tensor reduction techniques.

The modified phase-space integrals inherit both the UV divergences and
the IR singularities of the original loop amplitudes.  Dimensional
regularization and renormalization cannot be used without modifying
phase space, which impedes a straightforward interpretation of
physical events, so we do not consider it.  In the next two sections we
will instead introduce subtraction graphs as counterterms for the
elimination of UV divergences and discuss the treatment of infrared
poles.

\subsubsection{The role of multiple cuts}

The first line of (\ref{ftt}) can also be interpreted as the result of
the $k_0$ integration of the original loop, when the contour is closed
in the lower half plane and only single, simple poles are picked
up. Setting $i\epsilon$ to zero afterwards leads to wrong results if
poles fall together and form double or multiple poles. In (\ref{ftt}),
there are additional subleading contributions which are collected in
the sum in the second line, indicated by the dots in
(\ref{looptotree}). These terms give a non-vanishing contribution, if
the momenta of the propagators they were replaced with, go on-shell
simultaneously. Since after the $k^0$ integration there are still
$\delta$-functions left, these terms will get support for two
dimensional surfaces, lines or points in the three-dimensional phase
space volume, depending on the initial number of $\Delta_i$
functions. Since each $\Delta_i$ effectively lowers the dimension of
the integration by one, the contribution of these terms can be
calculated rather easily.

Replacing a propagator by a delta function reduces the number of
factors $i$ by one. Terms in (\ref{ftt}) with an even number of
$\Delta_i$ will therefore give an imaginary contribution to the final
result, terms with an odd number a real contribution.

Whether these terms give a non-vanishing contribution, can already be
inferred from the integrand of the terms in the first line of
(\ref{ftt}). When replacing the original integral by a sum over tree
graphs with one additional on-shell particle, the remaining
propagators can become singular, or in other words, internal lines can
get on-shell at certain values of the momentum $\vec{k}$. This leads
to a peak structure which is resembled by the sub-leading terms in
(\ref{ftt}). In section \ref{s_threshold_singularities} we will give
the construction of smoothing functions which cancel these peaks.

\subsection{Renormalization and Regularization}\label{s_ren_reg}

Going from Born-level to loop-level calculations, the initial relation
between the bare parameters in the Lagrangian and the physical ones is
destroyed. To restore this relation, renormalization constants are
introduced, which also absorb ultraviolet divergences arising in loop
calculations. These additional free parameters of the theory have to
be fixed by imposing renormalization conditions. 

As long as we are dealing with massive theories, a convenient set of
conditions is the on-shell renormalization scheme,
\begin{align}\label{ren_cond}
\left. \mbox{Re}\,\, i\Gamma^{(2)}_{\alpha\beta}(-p,p)
  \Phi^{\beta}(p)\right|_{p^2=m^2} &= 0 
&
\left.\Gamma^{(3)}(p_i,\lambda)\right|_{p_i^2=m^2} &= \lambda^3_0 
\nonumber\\
\mbox{Res}\left(-\Gamma^{(2)}(p)\right)^{-1}_{\fmslash p=m,p^2=m^2}&=1
&
\left.\Gamma^{(4)}(p_i,\lambda)\right|_{p_i^2=m^2} &= \lambda^4_0,
\end{align}
\noindent This requires the pole of the real part of propagators to
coincide with the corresponding particle mass, with residue~$1$.
Vertex functions are set equal to the tree level vertex functions for
on-shell external legs.

In massless QCD, there is no on-shell scheme, and it is important to
satisfy Ward (or Slavnov-Taylor) identities order by order.  However,
we can nevertheless impose physical renormalization conditions for
observable quantities and use them for defining appropriate
subtractions, and it is well known that Ward identities can be
satisfied, in any subtraction scheme, by introducing a complete set of
counterterms with fixed coefficients.  Since our proof-of-concept
example is in the context of massive QED, we postpone the construction
of QCD (and Standard-Model) counterterms to future work.

In multiplicative renormalization, the conditions (\ref{ren_cond}) are
used to relate and fix all of the renormalization constants $\delta
Z_i$. By eliminating all redundancies from the renormalization
conditions, one ends up with only a handful of constants with rather
simple analytic expressions. These are sufficient to renormalize any
diagrams in perturbation theory.

In the majority of one-loop calculations, ultraviolet divergent loop
integrals are evaluated in dimensional regularization or variations of
that scheme. Divergent pieces are extracted as analytic poles in the
regulator $\epsilon$ and cancelled against the poles in the
counterterms. Using the Feynman Tree Theorem to get a numerically
integrable expression for the loop integral, we do not consider an
extension of the dimension of integration and an algebraic reduction
to finally extract the ultraviolet divergent pieces. In contrast to
this procedure, we make use of a variation of the BPHZ procedure
\citerange{Bogoliubov:1957gp,Zimmermann:1969jj}, which results
in loop graphs acting as counterterms, which can be evaluated under
the same phase space integral over the loop momentum.

Consider a 1PI one-loop graph $\Gamma^n(p_1,\dots,p_n)$ with
superficial degree of divergence $\omega(\Gamma)$. We define the T
operator as a Taylor expansion around on-shell momenta $\bar{p}_i$,
with $\bar{p}_i^2=m_i^2$:
\begin{eqnarray}\label{t_op}
T\circ \Gamma^n(p_1,\dots ,p_n) &=&  \Gamma^n(\bar{p}_1,\dots ,\bar{p}_n)+
\sum\limits_{i}^{n-1}(p_i-\bar{p}_i)^\mu \left. \frac{\partial \Gamma^n}{\partial p_i^\mu}\right|_{p_1=\bar{p}_1,\dots ,p_n=\bar{p}_n}+\nonumber\\&&
\dots +\\ &&
\frac{1}{d!}\sum\limits_{i_1,\dots ,i_d}^{n-1} (p_{i_1}-\bar{p}_{i_1})^{\mu_1} \dots (p_{i_d}-\bar{p}_{i_d})^{\mu_d}\left.\frac{\partial^d \Gamma^n}{\partial p_{i_1}^{\mu_1}\dots\partial p_{i_d}^{\mu_d}}\right|_{p_1=\bar{p}_1,\dots ,p_n=\bar{p}_n}\nonumber,
\end{eqnarray}
\noindent up to $d=\omega(\Gamma)$. With this T operator, the renormalized 1PI n-point functions
\begin{equation}
\hat{\Gamma}^n(p_1,\dots ,p_n) =  \Gamma^n(p_1,\dots ,p_n) - T \circ \Gamma^n(p_1,\dots ,p_n)
\end{equation}
\noindent fulfill the renormalization conditions
(\ref{ren_cond}). Note, that we consider the external four-vectors in
Minkowski space and not Euclidean space, which is commonly used in
schemes derived from the BPHZ prescription. In our formulation, the
subtraction terms arising from (\ref{t_op}) can be complex
valued. However, only the real part is needed for renormalization. For
two-loop calculations, it therefore might be necessary to further
restrict the subtraction terms to be the real parts of the considered
graphs.

The expressions resulting from (\ref{t_op}) can also be interpreted as
loop graphs and are easily derived from the original Feynman graph. 

The hard scattering part of the virtual cross section in a NLO
computation of a $2\rightarrow n$ process can therefore be written as
\begin{equation}
\sigma^{(1)}_v \propto \int \! d\Pi_n \, 2\,\mbox{Re}({\mathcal{M}^\textindex{Born}}(\mathcal{M}^\textindex{loop}_n
+\mathcal{M}^\textindex{loop}_{n,\textindex{CT}})^*),
\end{equation}
\noindent where we left out any parton distribution function,
fragmentation or cut functions. We can now use the Feynman Tree
Theorem again to cut the loop graphs and collect the individual loop
integrals in a common phase-space integral:
\begin{equation}\label{sigma_virt}
\sigma^{(1)}_v \propto \!\!\int\! d\Pi_n \!\int\! \frac{d^3k}{(2\pi)^3} 2\mbox{Re}(\mathcal{M}^\textindex{Born}_n(\mathcal{M}^\textindex{Tree}_{n+1}+
\mathcal{M}^\textindex{Tree}_{n+1,\textindex{CT}})^*).
\end{equation}
\noindent The resulting integrals are ultraviolet finite by
construction. There are still ambiguities in (\ref{t_op}) for the
choice of the on-shell momenta $\bar{p}_i$. To assure local cancellation of
infrared divergences between the virtual matrix elements and the real
emission graphs, these have to be chosen in a specific way, which will
be described in section \ref{s_uv/ir}.

\subsection{IR cancellations}\label{s_ir}
Physically, an initial or final particle state cannot be distinguished
from a state with an additional number of soft photons. It was pointed
out by Kinoshita \cite{Kinoshita:1962ur}, Lee and Nauenberg
\cite{Lee:1964is} that the sum over all degenerate states is infrared
safe in each order of perturbation theory. In the case of QED
corrections this means that if the Born cross section of the real
emission process $\sigma^{(1)}_\textindex{re}$ with $n$ particles and
an additional photon in the final state is added to the virtual cross
section $\sigma^{(1)}_v$ the resulting total cross section is finite
\cite{Bloch:1937pw}. Assuming that the detector cannot resolve photons
of energy less than $\Delta E_s$, one can split up the real emission
cross section, which is of the same order in perturbation theory than
the virtual cross section, into a soft and a hard part,
\begin{eqnarray}
\sigma^{(1)}_\textindex{re} &=& \sigma^{(1)}_\textindex{soft}(\Delta E_s) + \sigma^{(1)}_\textindex{hard}(\Delta E_s)\\
\sigma^{(1)}_\textindex{soft}(\Delta E_s) &\propto&  \int d \Pi_n \! \int\limits^{\Delta E_s}\!\! \frac{d^3k}{(2\pi)^3 2E_k}\, |\mathcal{M}^\textindex{Born}_{n+\gamma}|^2\label{sigma_soft}\\
\sigma^{(1)}_\textindex{hard}(\Delta E_s) &\propto&  \hspace{-5mm} \int\limits_{E_k \ge \Delta E_s} \hspace{-4mm} d \Pi_{n+\gamma}\, |\mathcal{M}^\textindex{Born}_{n+\gamma}|^2
\end{eqnarray}
\noindent and add the soft part to $\sigma^{(1)}_{v}$. The form of the
two phase space integrals in (\ref{sigma_soft}) and (\ref{sigma_virt})
look very similar, they only differ by an implicit delta function
conserving overall momentum of the virtual and the real emission
process. We can therefore find a simple approximation of the real
emission processes for momenta $k<\Delta E_s$, such that the soft
infrared poles in the virtual graphs are cancelled locally on the
integrand level, allowing a numerical evaluation of
(\ref{sigma_virt}).
\begin{figure}[t]
\begin{center}
\begin{picture}(130,32)
\put(0,0){
\begin{minipage}[t]{13cm}
\includegraphics[width=13cm]{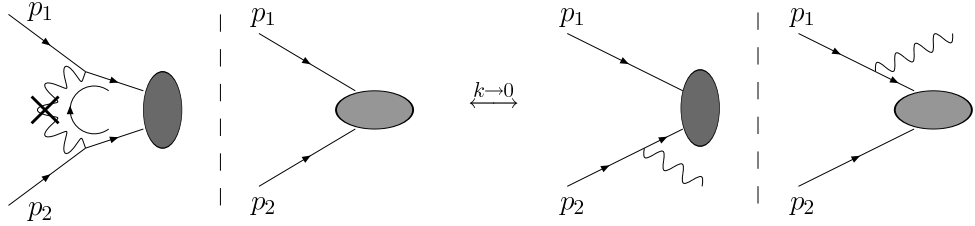}
\end{minipage}}
\end{picture}
\end{center}
\caption{Infrared divergent contribution of a loop-Born interference
  term: In the limit $|\vec{k}|\rightarrow 0$, the loop integral with
  the cut photon line diverges. This divergence is compensated by the
  product of the two corresponding real emission
  diagrams.} \label{f_ir_div} \vspace{3mm}
\end{figure}

Using the Tree Theorem to cut a loop graph, the infrared divergent
part of the resulting integrand arises solely from the cut of massless
propagators connecting two external on-shell particles. As an example,
consider the photon exchange of two charged incoming fermions, as
shown in figure \ref{f_ir_div}. Before cutting any propagator the
relevant part of the integrand reads:
\begin{equation}
\frac{d^4k}{(2\pi)^4} \frac{\ldots(\fmslash k+ \fmslash p_1+m_1)\gamma^\mu u_\lambda(p_1)}{k^2+2kp_1}\cdot\frac{-ig_{\mu\nu}}{k^2}\cdot\frac{\ldots(-\fmslash k+ \fmslash p_2+m_2)\gamma^\nu u_\kappa(p_2)}{k^2-2kp_2},
\end{equation}
\noindent where for better readability we chose the integration
momentum such that it is equivalent to the momentum flowing through
the photon line. Cutting this line, we get
\begin{equation}
\frac{d^3k}{(2\pi)^3 2|\vec{k}|}\!\!\sum\limits_\sigma\!\! \left. \frac{\ldots(\fmslash k+ \fmslash p_1+m_1)\gamma^\mu u_\lambda(p_1)}{2kp_1}\!\cdot\! \epsilon_\mu(k,\sigma)\epsilon^*_\nu(k;\sigma)\!\cdot\!\frac{\ldots(-\fmslash k+ \fmslash p_2+m_2)\gamma^\nu u_\kappa(p_2)}{-2kp_2}\right|_{k_0=|\vec{k}|},
\end{equation}
\noindent which is logarithmically divergent in the limit
$|\vec{k}|\rightarrow 0$. This loop amplitude is multiplied with a
Born amplitude ${\mathcal{M}_\textindex{Born}}^\dagger$. Neglecting
the term $\fmslash k$ in the numerator we can shift the incoming
photon line on the leg of particle $1$ to the Born matrix element,
whose relevant part is just the Dirac wave function $\bar{u}_\lambda$
of this particle connected to a Dirac matrix $\gamma^\rho$:
\begin{equation}
\frac{\ldots(\fmslash p_1+m_1)\gamma^\mu u_\lambda(p_1) \bar{u}_\lambda(p_1)\gamma^\rho\ldots}{2kp_1}\cdot \epsilon_\mu(k,\sigma)\rightarrow
\frac{\ldots u_\lambda(p_1) \bar{u}_\lambda(p_1)\gamma^\mu (\fmslash p_1+m_1) \gamma^\rho\ldots}{2kp_1}\cdot \epsilon_\mu(k,\sigma).
\end{equation}
Taking the hermitian conjugate of the right part and adding another $\fmslash k$ term in the numerator, we get
\begin{equation}
\cdots\frac{\bar{u}_\lambda(p_1)\gamma^\mu (\fmslash p_1+m_1) \gamma^\rho\ldots}{2kp_1}\cdot \epsilon_\mu(k,\sigma)\rightarrow
\frac{\ldots\gamma^\rho(-\fmslash k+\fmslash p_1+m_1)  \gamma^\mu u_\lambda(p_1)}{2kp_1}\epsilon^*_\mu(k,\sigma).
\end{equation}
This corresponds to the Born diagram with an additional outgoing
on-shell photon attached to one external line times an overall factor
$-1$, which would usually arise from the propagator adjacent to the
emitted photon,
\begin{equation}
(k-p_1)^2-m^2=-2kp_1.
\end{equation}
After the above transformations, which involved manipulations of the
numerator of $\mathcal{O}(|\vec{k}|)$, the original loop becomes also
a Born graph with a photon emitted from particle $2$. Thus, in the
limit $|\vec{k}|\rightarrow 0$ the divergent piece of the one-loop
contribution is exactly cancelled by the product of two real emission
diagrams.

If we obtain the two real emission diagrams from the loop diagram in
the way shown above, momentum conservation is violated in both graphs
at some vertex. This happens, because the initial delta function
conserving the momenta of the external particles in the initial loop
diagram is still present in the real emission diagrams. The additional
particle violates momentum conservation. Although the integration
measure is the same in both cases, the true real emission diagrams are
accompanied by a delta function $\delta (P-\sum q_f-q_\gamma )$
conserving overall momentum.

Momentum conservation at some vertices is also violated in the soft
photon approximation, e.g. cf
\cite{Yennie:1961ad,Denner:1991kt}. Here, following the same reasoning
as above, the contribution of real emission diagrams in the soft limit
is approximated by the Born amplitude times a prefactor. This factor
can be evaluated analytically, e.g., when regulated by a photon
mass. Adding this expression to the analytic results of loop graphs,
the divergent terms, logarithms of the photon mass, cancel.

If we want to evaluate the soft real emission diagrams and the loop
corrections under the same integral, we need a method of implementing
the projection of the, e.g. $2\rightarrow n +\gamma$ real emission
graphs onto the $2\rightarrow n$ virtual graphs. Reversing the above
approximation, we could add the product of two real emission diagrams
with momentum violation at the first vertex after the emission of the
massless particle. However, adding the product of the two diagrams
will only cancel the infrared divergence in the product of the loop
graph with the corresponding Born graph. We intend to apply the
Feynman Tree Theorem on the amplitude level and compute the interference
with the Born terms after summation of all contributing loop
graphs. Applying the Tree Theorem to products of loop and Born graphs
would lead to a drastic increase of the number of terms, if a process
with several Born terms is considered.

We therefore need a prescription of incorporating the effect of real
emission diagrams in single infrared divergent loop graphs. In the
following, we set the term with the cut propagator associated with the
massless particles to zero for $|\vec{k}|<\Delta E_s$.  This simple
prescription is sufficient for the QED process considered in the
present paper, as demonstrated by the numerical results.  For a
generic solution one would have to adapt a method such as phase-space
slicing or dipole subtraction to the present situation.

\subsection{UV subtractions and IR divergences}\label{s_uv/ir}

There is a one-to-one correspondence between the interference terms of
Born graphs with loop diagrams with a cut massless particle and the
product of two real emission diagrams of this particle, regarding the
infrared behavior. This means that for any virtual infrared divergence
in an unrenormalized loop, there is a product of two real emission
diagrams with emission of this particle from external legs, cancelling
this divergence. 

However, there are still infrared divergent terms left. On the one
hand those which correspond to the square of one real emission diagram
and therefore would arise from self-energy corrections to an external
particle, which are set to zero by the on-shell renormalization
scheme. On the other hand, further infrared divergent contributions
arise from subtraction diagrams used as counterterms to cancel
ultraviolet divergent virtual contributions. In this section we will
argue that by a certain choice of the subtraction diagrams these
additional infrared divergences cancel and the final result is
infrared finite.

The following considerations apply to the case of QED, but the
generalization to QCD and the SM, given a suitable renormalization
scheme, is straightforward.

In QED, there are three primitively ultraviolet divergent graphs
$\Gamma$. These are the photon and electron self-energy and the vertex
correction. The renormalized one-loop photon self-energy is infrared
finite. In the following we will argue that the infrared divergent
terms of the square of real emission diagrams and of the electron
self-energy will be compensated by the subtraction diagrams of the
vertex corrections.

Consider an amplitude $\mathcal{M}_0$ with an incoming on-shell
electron with momentum $p$ and mass $m$. The electron line is attached
to a vertex $\mathcal{V}$. If we radiate off a real photon from this
line and square the amplitude, the infrared divergent term can be
obtained from soft photon approximation and reads:
\begin{equation}\label{ire}
I_{re}= -e^2 |\mathcal{M}_0|^2\cdot\!\!\int\limits^{\Delta E_s}\!\!\frac{d^3k}{(2\pi)^3 2|\vec{k}|}\frac{m^2}{(pk)^2}.
\end{equation}
The radiative correction to the vertex $\mathcal{V}$ is infrared
divergent if the outgoing fermion line is on-shell. As shown in
section \ref{s_ir}, this divergence is cancelled by the product of two
real emission diagrams with the photon attached to the incoming and
outgoing fermion line, respectively. As argued in section
\ref{s_ren_reg}, to relate experimental results with theoretical
calculations we subtract the vertex correction at the Thomson limit
where the momentum of the photon attached to the vertex $\mathcal{V}$
is zero and the fermion going through the vertex is on-shell. If we
split the subtraction term in two pieces, where in one term the
on-shell fermion line through the vertex $\mathcal{V}$ has momentum
of the incoming electron and in the second term it has the momentum of
the outgoing fermion, the loop contribution of one of these terms is
\begin{equation}\label{vertex_loop}
\frac{1}{2}(-ie^2)\int\frac{d^4k}{(2\pi)^4}\cdots\frac{\gamma_\alpha(\fmslash p+\fmslash k+m)(-ie\gamma_\mu)
(\fmslash p+\fmslash k+m)\gamma^\alpha}{k^2(k^2+2pk)^2}\cdots,
\end{equation}
\noindent where we explicitly pulled out the couplings $(-ie)$ from
the $\gamma_\alpha$ and the factors $(i,-i)$ coming from the
propagators. As can be seen from (\ref{vertex_loop}), the ultraviolet
contribution of the subtraction terms are independent from the
external momenta of the vertex correction. Thus, the subtraction graph
can be split in parts with different momentum assignments without
affecting the cancellation of the UV divergence.

Cutting the photon line of the loop with momentum $k$ and neglecting
terms proportional to $k$ in the numerator we get:
\begin{equation}
-\frac{e^2}{2}\int\frac{d^3k}{(2\pi)^3 2|\vec{k}|}\cdots\frac{4m(-iep_\mu)}{(2pk)^2}\cdots.
\end{equation}
If we {\it straighten} the fermion line such that throughout the graph
it has the on-shell momentum $p$, with no momentum flowing in or
out at any vertex,  this fermion line can be written as a chain of products
of Dirac wave functions and $\gamma$-matrices:
\begin{equation}
\ldots \bar{u}(p)\gamma_\kappa u(p)\bar{u}(p) \gamma_\lambda u(p)\bar{u}(p) p_\mu u(p)\bar{u}(p)\ldots,
\end{equation}
\noindent where at the place of the former vertex correction only a
factor proportional to $p_\mu$ remains. Making use of the Gordon
identity
\begin{equation}\label{gordon}
\bar{u}(p) p_\mu u(p) = m \bar{u}(p) \gamma_\mu u(p),
\end{equation}
the infrared divergent term can be factored out of the amplitude.
\begin{equation}
-\frac{e^2}{2}\int\frac{d^3k}{(2\pi)^3 2|\vec{k}|}\frac{4m^2}{(2pk)^2}\cdot \mathcal{M}'_0.
\end{equation}
Here, $\mathcal{M}'_0$ is the matrix element $\mathcal{M}_0$ with all
lines attached to the fermion line bearing zero momentum. Since we
factored out the infrared divergent part, we can divide by
$\mathcal{M}'_0$ and multiply by $\mathcal{M}_0$, to compensate for
the projection onto the straight fermion line. In explicit
calculations we therefore subtract the graph with the vertex
correction, where we keep the momentum of the fermion
fixed throughout the graph and neglect the denominator of the rest 
of the matrix element which does not belong to the loop.
We then divide by the same matrix element without vertex
correction and multiply by the basic amplitude $\mathcal{M}_0$. Doing
so, we apply the correct subtraction terms to the unrenormalized loop
graph in the sense that in the limit, where the momentum of the photon
attached to $\mathcal{V}$ vanishes, the one-loop graph and the
subtraction graphs cancel each other. This leaves the Born graph,
which just contains the electric charge at the vertex,
$(-ie\gamma_\mu)$, as required by the renormalization
conditions. Since the ultraviolet contributions of the subtraction
graphs are independent of the external momenta ab initio, this
procedure does not invalidate the UV cancellation.

The resulting infrared divergent term of the interference with the
amplitude $\mathcal{M}_0$, which comes with a factor $2$ in the final
cross section, becomes:
\begin{equation}\label{iv1}
I_{v_1}=-e^2 |\mathcal{M}_0|^2 \cdot \int\frac{d^3k}{(2\pi)^3 2|\vec{k}|}\frac{m^2}{(pk)^2},
\end{equation}
\begin{figure}[t]
\begin{center}
\begin{picture}(148,20)
\put(0,0){
\begin{minipage}[t]{14.8cm}
\includegraphics[width=14.8cm]{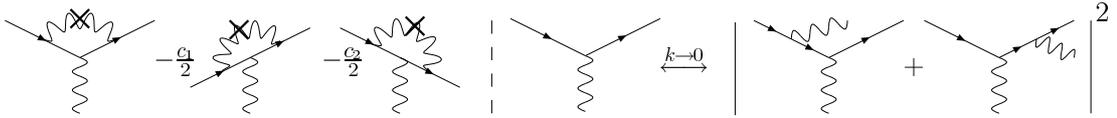}
\end{minipage}}
\end{picture}
\end{center}
\caption{UV-IR Subtraction Terms: The vertex correction is renormalized by the subtraction of two graphs with zero  incoming photon momentum. When the photon propagator of the loops is cut, the arising infrared divergences in the interference term are compensated by the product of real emission diagrams depicted on the right hand side.} \label{f_uv_ir}
\end{figure}

\noindent which exactly cancels the above infrared divergent term
(\ref{ire}) of the soft real emission when subtracted from the
unrenormalized vertex correction. The second half of the subtraction
term, $I_{v_2}$, is equivalent to $I_{v_1}$ with $p$ replaced by the
on-shell momentum $\bar{q}$ of the outgoing fermion. If this is an
external particle, the infrared divergence is cancelled by the
corresponding real emission diagrams and we are finished. This case is
summarized in figure \ref{f_uv_ir}. On the left hand side we depicted
the renormalized vertex correction. The coefficients are given by
$c_i=\mathcal{M}_0\mathcal{M}_0^{'-1}(p_i)$. When the photon lines of
the loops are cut, the infrared divergences arising in the
interference term are cancelled by the real emission diagrams shown on
the right hand side.

If the outgoing fermion belongs to an internal line, we will show in
the following that the infrared divergent part is cancelled by the
subtraction terms to the self-energy correction to this internal line.

As was shown in section \ref{s_ren_reg}, in the case of an self-energy
correction to an internal charged fermion line, we need two
subtraction terms to cancel the UV divergence. We subtract the same
graph at an on-shell momentum $\bar{q}$ aligned to the original
momentum $q$ and the derivative of the self-energy with respect to $q$
at $\bar{q}$. The relevant part of this diagram is given by the
electron self-energy
\begin{equation}
-e^2\cdot\int \!\!\frac{d^4k}{(2\pi)^4}\cdots\frac{\gamma_\alpha(\fmslash q +\fmslash k +m)\gamma^\alpha}{k^2((k+q)^2-m^2)}\cdots.
\end{equation}
Replacing $q$ by $\bar{q}$ will not lead to an infrared singular term, since the singularity in the denominator is cancelled by the integration measure. Making use of the identity
\begin{equation}
\frac{\partial}{\partial q^\mu}\frac{i}{(\fmslash k + \fmslash q -m)}=\frac{i}{(\fmslash k + \fmslash q -m)}
i\gamma_\mu\frac{i}{(\fmslash k + \fmslash q -m)},
\end{equation}
\noindent the second subtraction term can simply be obtained by straightening the fermion line through the self-energy part, insertion of a Dirac gamma matrix $\gamma_\mu$ in the fermion line and multiplying by $(q-\bar{q})^\mu$:
\begin{equation}
-ie^2\cdot \int \!\!\frac{d^4k}{(2\pi)^4}\cdots\frac{\gamma_\alpha(\fmslash {\bar{q}} +\fmslash k +m)i(\fmslash q -\fmslash {\bar{q}})(\fmslash {\bar{q}} +\fmslash k +m)\gamma^\alpha}{k^2((k+\bar{q})^2-m^2)^2}\cdots.
\end{equation}
Cutting the photon line and simplifying the numerator, the divergent part is
\begin{equation}
-e^2\cdot \int \!\!\frac{d^3k}{(2\pi)^3 2|\vec{k}|}\frac{i(q-\bar{q})^{\mu}\bar{q}_\mu}{(\bar{q}k)^2}\cdots(-\fmslash {\bar{q}}+2m)\cdots.
\end{equation}
We can again interpret the fermion line as a chain of Dirac wave
functions. Thus, the factor $(-\fmslash{\bar{q}}+2m)$ simplifies to
$m$ and with the use of the Gordon identity (\ref{gordon}) we write
\begin{equation}
 i(q-\bar{q})^{\mu}\bar{q}_\mu=m\cdot i(q-\bar{q})^{\mu}\gamma_\mu = m\cdot i(\fmslash q -m).
\end{equation}
\noindent This is $m$ times $\mathcal{M}_0'$, the matrix element with
an insertion of $i(\fmslash q -m)$ at the place of the original
self-energy correction. Like in the case of the vertex correction we
then have a Lorentz scalar factored out of the amplitude
$\mathcal{M}_0'$ with a straight fermion line. To project this
counterterm on the onto the matrix element with the original
kinematics, $q$ instead of $\bar{q}$, we multiply by
$\mathcal{M}_0^{'-1}$ and the Born matrix element with the additional
insertion $i(\fmslash q -m)$ at the place of the original self-energy
correction. With the two propagators connecting to this 2-point vertex
we have:
\begin{equation}
\cdots \frac{i(\fmslash q + m)}{q^2-m^2}i(\fmslash q -m)\frac{i(\fmslash q + m)}{q^2-m^2}\cdots=\cdots\frac{-i(\fmslash q + m)}{q^2-m^2}\cdots,
\end{equation}
\noindent which is proportional to the Born matrix element
$\mathcal{M}_0$. Multiplying again with $2 \mathcal{M}_0$ to calculate
the interference term contributing to the cross section, the infrared
divergent part of the subtraction terms to the electron self-energy
reads:
\begin{equation}\label{iee}
I_{\Sigma}= 2 e^2 |\mathcal{M}_0|^2\cdot \int \!\!\frac{d^3k}{(2\pi)^3 2|\vec{k}|}\frac{m^2}{(\bar{q}k)^2}.
\end{equation}
When subtracted from the unrenormalized self-energy correction, half
of the infrared divergent term cancels the contribution coming form
the subtraction graph of the vertex correction $I_{v_2}$, obtained
from (\ref{iv1}) by replacing $p$ with $\bar{q}$. Following the
fermion line further we again come to a vertex and its correction
terms will cancel the second half of the infrared divergent term of
the subtraction graphs of electron self-energy. This goes on until the
last vertex, where the second infrared contribution of its subtraction
graphs will then be compensated by the squared amplitude of the real
emission of a photon of the external line following this vertex.

This completes our renormalization prescription, which trivially
extends to processes with multiple Born graphs. We constructed
subtraction graphs which cancel all possible ultraviolet divergences
and give further infrared divergent contributions such that in the sum
of all graphs contributing to a given process the infrared divergences
cancel. Furthermore the renormalized vertex functions obey the
renormalization conditions (\ref{ren_cond}) such that the
experimentally measured observables can directly be related to the
theoretical predictions without any further analytic correction.

We showed that in QED we have a complete prescription to incorporate
the on-shell renormalization scheme and cancel all infrared and
ultraviolet divergences. Although not rigorously proven, this method
should also be applicable to the electroweak standard
model. Subtraction graphs to only ultraviolet divergent vertices can
be found by the introduced BPHZ mechanism. In case of a photonic
correction, where infrared divergences are expected, the proposed
method of this section should also lead to finite
results. 

\subsection{Threshold singularities}\label{s_threshold_singularities}
When the momentum integration is performed in the first line of
(\ref{ftt}), the integrand might get peaks in parts of the phase space
where momenta of un-cut propagators are on-shell. These regions are
open or closed two-dimensional surfaces in the three-dimensional
integrand. Intersections of these surfaces correspond to kinematic
situations where two or more momenta of internal lines become on-shell
at the same time. The occurrence of such peaks, although analytically
integrable, leads to problems in the numerical evaluation of the
integrand. Subtraction terms with zero real value but with the same
peak structure will smooth the integrand and allow for a better
convergence in the numerical evaluation. In this section we will give
the conditions under which peaks of the integrand arise and calculate
the corresponding fixing functions.

Cutting a propagator $P_i$ in a loop leads to a delta function
$\Delta^l_i$, given in (\ref{Deltal}), which effectively sets the
four-vector $k+p_i$ on its mass-shell at $(k+p_i)^2=m_i^2$ with
positive zero component. There are two possible situations under which
another propagator $P_j$ can get singular.  Its
original $k_0$-poles lie either in the lower or upper half plane, which
corresponds to the positive and negative zero components $\pm
E_j$. After the $k_0$ integration the relevant term of the integrand
in the Tree Theorem (\ref{ftt}) reads:
\begin{equation}\label{DPR}
\Delta_i^l P_j R(k)=\frac{1}{2E_i}\frac{i}{[(p_j^0-p_i^0)+(E_i-E_j)][(p_j^0-p_i^0)+(E_i+E_j)]}R(-p_i^0+E_i,\vec{k}),
\end{equation}
\noindent where $R(k)$ is the analytic remainder of the term,
containing the numerator and denominators of further propagators which
for now are assumed to be non-singular in the integration volume. If
the first factor in the denominator vanishes for some $\vec{k}$, both
momenta of the two propagators $P_i$ and $P_j$ get on-shell with a
positive zero component. In other words, at this constellation of the
integration momentum $\vec{k}$, the two poles in the lower $k_0$ half
plane of the original loop momentum coincide. Therefore, we will also
encounter this singularity when propagator $P_j$ is cut:
\begin{equation}
\Delta_j^l P_i R(k)=\frac{1}{2E_j}\frac{i}{[(p_i^0-p_j^0)+(E_j-E_i)][(p_i^0-p_j^0)+(E_j+E_i)]}R(-p_j^0+E_j,\vec{k}).
\end{equation}
In the limit of the first factor becoming zero, the residue of the
combined contribution vanishes:
\begin{eqnarray}
\lim_{(p_j^0-p_i^0)+(E_i-E_j)\rightarrow 0}\!\!\!\!&&\left((p_j^0-p_i^0)+(E_i-E_j)\right)\left(\Delta_i^l P_j R(k)+\Delta_j^l P_i R(k)\right)\nonumber\\
&&=\frac{i}{2E_i2E_j}\left(R(-p_i^0+E_i,\vec{k})-R(-p_i^0+E_i,\vec{k})\right)=0.
\end{eqnarray}
Thus, if the two terms are added, the peaks will compensate each other
and the integrand can safely be evaluated numerically in this case.

Whenever a propagator $P_j$ gets singular in the integration region,
one of the two corresponding delta functions, $\Delta^l_j$ or
$\Delta^u_j$ get support at these regions in phase space. Therefore,
the final result for the loop integral gets further contributions from
the higher order terms in the Feynman Tree Theorem (\ref{ftt}), if
these threshold peaks in the integration region are encountered during
the integration of the leading order terms in (\ref{ftt}).

The situation described above would correspond to a coincidence of two
poles in the lower $k^0$ half plane. However, in the sum of the tree
graphs, there are two peaks arising from two different tree graphs,
which cancel in the sum. Accordingly, we also did not get a term in
(\ref{ftt}) with only two $\Delta^l$. Therefore, just like there is no
peak in the sum of the tree level contribution there is also no
imaginary contribution to the final result in this case.

In the case where the second factor in (\ref{DPR}) becomes singular,
we will get a peak which remains in the sum of the tree graphs. This
happens, if one pole of the lower $k_0$-half plane coincides with a
pole in the upper half plane. In this case, there is a term with one
$\Delta^l$ and one $\Delta^u$ which get support simultaneously and
lead to the imaginary part of the loop graph.

\subsubsection{Conditions for Internal Singularities}\label{s_cond_int}
\begin{figure}[t]
\begin{center}
\begin{picture}(80,60)

\put(0,0){\begin{minipage}[t]{7cm}
\includegraphics[width=7cm]{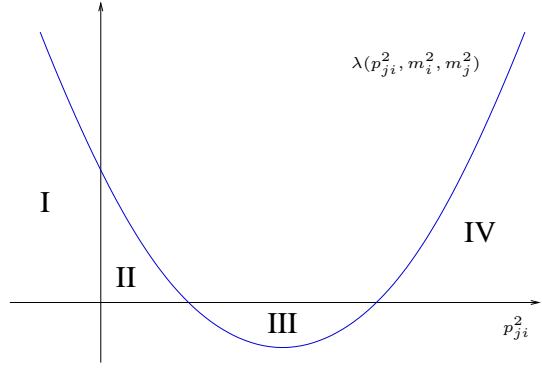}
\end{minipage}}
\put(45,55){\mbox{\tiny $\lambda(p_{ji}^2,m_i^2,m_j^2)$}}
\put(65,20){\mbox{\tiny $p_{ji}^2$}}
\end{picture}
\vspace{-1.7cm}
\caption{Different regions for $p_{ji}^2$, defined by the origin and the two zeros of the kinematical function $\lambda$. The occurrence and the structure of singularities of the integrand depend on the value of $p_{ji}^2$.} \label{f_lambda}  
\end{center}
\end{figure}
We can look for general conditions under which a loop propagator $P_j$
becomes singular if we cut a propagator $P_i$, and give equations for
the corresponding surfaces. Cutting propagator $P_i$, we have:
\begin{eqnarray}
0&\cond&\left.(k+p_j)^2-m_j^2\right|_{k^0=-p_i^0+\sqrt{(\vec{k}+\vec{p}_i)^2+m_i^2}},\nonumber\\
&\cond&\left.((k+p_i)+(p_j-p_i))^2-m_j^2\right|_{k^0=-p_i^0+\sqrt{(\vec{k}+\vec{p}_i)^2+m_i^2}},\nonumber\\
&\cond&\left. m_i^2-m_j^2+2(k+p_i)(p_j-p_i)+(p_j-p_i)^2
\right|_{k^0=-p_i^0+\sqrt{(\vec{k}+\vec{p}_i)^2+m_i^2}}.\label{cond0}
\end{eqnarray}
The occurrence and the shape of the peaks depend on the value of
$p^2_{ji}\equiv(p_j-p_i)^2$. We can distinguish $4$ kinematic regions,
separated by $p^2_{ji}=0$ and the two nodes
$\lambda(p_{ji}^2,m_i^2,m_j^2)=0$ of the kinematical function $\lambda$,
defined by
\begin{equation}\label{lambda_kin}
\lambda(x,y,z)=x^2+y^2+z^2-2xy-2xz-2yz.
\end{equation}
The different regimes are depicted in figure \ref{f_lambda}.

We switch to Lorentz frames where the calculation of solutions to
(\ref{cond0}) is particularly simple. We first discuss the case of
negative $p_{ji}^2$.
\newline

$\bullet$ $p_{ji}^2<0$, Region {\bf I}
\newline

\noindent Here, we cannot find a rest frame of $p_j-p_i$, however, we
can define a Lorentz transformation projecting $p_j-p_i$ onto the
z-axis:
\begin{eqnarray}
\Lambda^{\mu\nu}(p_j-p_i)_\nu &=& (0,0,0,p_{ji}^z),\\
-{p_{ji}^z}^2&=&(p_j-p_i)^2,\\
k^{\mu\prime} &\equiv &\Lambda^{\mu\nu}k_\nu.
\end{eqnarray}
Using this transformation we get for (\ref{cond0}):
\begin{eqnarray}
0&\cond & m_i^2-m_j^2+2(k+p_i)(p_j-p_i)+(p_j-p_i)^2, \nonumber\\
&=& m_i^2-m_j^2+2k(p_j-p_j)+p_j^2-p_i^2,\nonumber\\
&\stackrel{L.T.}{=}&m_i^2-m_j^2-2k'_zp_{ji}^z+p_j^2-p_i^2,\nonumber\\
\Longrightarrow k'_z&=&\frac{(p_j^2-m_j^2)-(p_i^2-m_i^2)}{2p_{ji}^z}.\label{kz}
\end{eqnarray}
There are no further conditions for this singularity to appear. Thus,
we will always encounter it in the case of $(p_j-p_i)^2<0$. However,
if we cut propagator $P_j$, propagator $P_i$ will exactly have the
same singularity with opposite sign. This can be seen from the
derivation of (\ref{kz}) by interchanging indices $i$ and $j$, but
applying the same Lorentz transformation as before. As a result we
have the situation described in the beginning of this section and the
two peaks will cancel each other in the sum of the tree graphs.
\newline

$\bullet$ $p_{ji}^2>0$, Regions {\bf II} - {\bf IV}
\newline

\noindent Here, we apply a Lorentz transformation to get into the rest
frame of $p_j-p_i$:
\begin{eqnarray}
(p_j-p_i)^\mu\rightarrow p^\mu_{ji}&=&\Lambda^{\mu\nu}(p_j-p_i)_\nu=(p_{ji}^0,\vec{0}),\\
{p_{ji}^0}^2&=&(p_j-p_i)^2.
\end{eqnarray}
We indicate $k'$ as the Lorentz transformed on-shell
momentum\footnote{Bold letters indicate absolute values of spatial
  vectors}:
\begin{equation}
k^{\mu\prime}=(\sqrt{\mathbf{k'}^2+m_i^2},\vec{k'})=\Lambda^{\mu\nu}(k+p_i)_\nu.
\end{equation}
With this parameterization of $k^{\mu\prime}$ the Lorentz
transformation has to be of proper orthochronous type, which implies
that $p_{ji}^0$ may be both positive or negative. We thus get from
(\ref{cond0}):
\begin{eqnarray}
&& m_i^2-m_j^2 + 2p^0_{ji}\sqrt{\mathbf{k'}_s^2+m_i^2}+{p^0_{ji}}^2=0, \label{propeq0}\\
&& \sqrt{\mathbf{k'}_s^2+m_i^2}=\frac{m_j^2-m_i^2-{p^0_{ji}}^2}{2p^0_{ji}},\label{poscond1}\\
\Longrightarrow \mathbf{k'}_s^2&=&\frac{1}{4{p^0_{ji}}^2}(m_i^4+m_j^4+
{p^0_{ji}}^4-2m_i^2{p^0_{ji}}^2-2m_j^2{p^0_{ji}}^2-2m_i^2m_j^2).\label{poscond2}
\end{eqnarray}
Therefore, the integrand gets singular at
\begin{equation}\label{ks}
\mathbf{k}_s=\frac{\lambda^{\frac{1}{2}}({p^0_{ji}}^2,m_i^2,m_j^2)}{2|p^0_{ji}|},
\end{equation}
\noindent the surface of a sphere with radius $\mathbf{k}_s$. The
kinematical function $\lambda$ is defined in (\ref{lambda_kin}). To get to
equation (\ref{ks}), two conditions have to be met to satisfy
equations (\ref{poscond1}) and (\ref{poscond2}):
\begin{eqnarray}
\frac{m_j^2-m_i^2-{p^0_{ji}}^2}{2p^0_{ji}}>0, \label{condI}\\
\lambda({p^0_{ji}}^2,m_i^2,m_j^2)>0.\label{condII}
\end{eqnarray}
The function $\lambda({p_{ji}^0}^2,m_i^2,m_j^2)$ is positive for
${p_{ji}^0}^2<(m_i-m_j)^2$ and ${p_{ji}^0}^2>(m_i+m_j)^2$. Condition
(\ref{condII}) therefore defines the three kinematic ranges for
$p_{ji}^2>0$, depicted in figure \ref{f_lambda}. If the kinematical
function $\lambda$ is positive we have to check if condition
(\ref{condI}) is fulfilled. We also check the similar condition of
propagator $P_i$ getting singular if propagator $P_j$ is cut. This
condition reads
\begin{equation}\label{condIII}
\frac{m_i^2-m_j^2-{p_{ji}^0}^2}{2p_{ji}^0}<0.
\end{equation}
\begin{itemize}

\item {\bf II}: $0<p_{ji}^2<(m_i-m_j)^2$

Suppose $m_i>m_j$. Then condition (\ref{condI}) simplifies to
$p_{ji}^0<0$. The numerator of (\ref{condIII}) is:
\begin{equation}
m_i^2-m_j^2-{p_{ji}^0}^2>m_i^2-m_j^2-(m_i^2-2m_im_j+m_j^2)=2m_j(m_i-m_j)>0.
\end{equation}
Thus, condition (\ref{condIII}) is now $p_{ji}^0<0$, the same as
(\ref{condI}). Therefore, if these conditions are fulfilled, we will
get a singularity in both terms, when propagator $P_i$ is cut and
propagator $P_j$ gets singular and vice versa. Again, this is the
situation described in the introduction to this section, when both
zero components of the on-shell momenta are positive. This can be seen
in the following. If propagator $P_i$ is cut, the denominator of
propagator $P_j$ can be written as:
\begin{equation}\label{rootproduct}
\left[\sqrt{\mathbf{k'}^2+m_i^2}+p_{ji}^0-\sqrt{\mathbf{k'}^2+m_j^2}\right]\cdot
\left[\sqrt{\mathbf{k'}^2+m_i^2}+p_{ji}^0+\sqrt{\mathbf{k'}^2+m_j^2}\right].
\end{equation}
The first factor corresponds to the pole in the lower $k_0$ half
plane. Its vanishing indicates the coincidence of the original poles
of the two propagators. We have
\begin{equation}
\sqrt{\mathbf{k'}^2+m_i^2}+p_{ji}^0=\frac{m_j^2-m_i^2+{p_{ji}^0}^2}{2p_{ji}^0}>\frac{m_j^2-2m_im_j+m_j^2}{2p_{ji}^0}=\frac{2m_j(m_j-m_i)}{2p_{ji}^0}>0.
\end{equation}
Therefore, the second factor is strictly positive and the singularity
can only arise from the first factor. Thus, the two poles of the lower
half plane fall together and the corresponding peaks cancel each other
as shown above. The analysis for $m_j>m_i$ is analogous and leads to
the same result.

\item {\bf III}: $(m_i-m_j)^2<p_{ji}^2<(m_i+m_j)^2$

In this region condition (\ref{condII}) is not fulfilled. Therefore
the integrand does not get a singular contribution from propagator
$j$.

\item {\bf IV}: $(m_i+m_j)^2<p_{ji}^2$

In this case the numerator of condition (\ref{condI}) as well of
condition (\ref{condIII}) is strictly negative. Thus, exactly one of
the two condition is fulfilled. Note also that if condition
(\ref{condI}) is fulfilled, meaning $p_{ji}^0<0$, the second factor of
(\ref{rootproduct}) vanishes. Thus, while one propagator gets on-shell
with a positive zero component, the other has a negative zero
component. Here, the singularity is not compensated by another term in
the Tree Theorem, and we will add a fixing function to cancel this
singularity, as will be shown in section \ref{s_fix_fct}. In this
kinematic range, the final result also gets a contribution from the
double delta terms in equation (\ref{ftt}). This adds to the imaginary
part of the integral. Here, we have exactly the situation of the
optical theorem, where to calculate the imaginary part of a forward
scattering amplitude two propagators are set on-shell simultaneously.
\end{itemize}

$\bullet$ $\lambda=0$
\newline

There are two more cases which we have not discussed yet:
$p_{ji}^2=(m_i-m_j)^2$ and $p_{ji}^2=(m_i+m_j)^2$. In case the two
points fall together, we either have $m_i=0$ or $m_j=0$ or both at the
same time. These are mass singularities, where the integrand is not
just singular but divergent. These divergences are compensated by the
addition of real emission graphs, as discussed in section
\ref{s_uv/ir}.

If the kinematical function $\lambda$ gets zero at two different
values of $p_{ji}^2$, which happens if both masses are non-zero, the
argumentation of the case $p_{ji}^2<(m_i-m_j)^2$ does not change in
the limit $p_{ji}^2=(m_i-m_j)^2$ and we do not encounter any peaks in
the integrand. However, if $p_{ji}^2=(m_i+m_j)^2$, we are at the
threshold where the two real particles with masses $m_i$ and $m_j$ can
be produced at the same time. These are Coulomb singularities, where
higher order corrections in perturbation theory can become equally
important. To get meaningful results, resummation methods can be
applied, e.g. cf. \cite{Fadin:1987wz}. In general, these peaks are
pointlike and integrable, setting the origin of the integration
variables to this point and using spherical coordinates will smooth
these peaks.

\subsubsection{Fixing Functions}\label{s_fix_fct}
We are now going to construct subtraction terms (which we denote as
\emph{fixing functions}) which will compensate the singularities of
the integrand without adding a real part to the result\footnote{The
  idea of adding a zero to the integrand which smooths the peaks is
  taken from \cite{Diplomarbeiten}. In this section we will give an
  construction of single fixing functions and also derive fixing
  functions in case of overlapping peaks.}.

Suppose we have replaced propagator $P_i$ with $\Delta^l_i$ and
propagator $P_j$ gets singular on a surface in the integration volume,
thus conditions (\ref{condI}) and (\ref{condII}) are fulfilled. The
integral is then
\begin{equation}
\int \frac{d^3k}{2\sqrt{(\vec{k}+\vec{p}_i)^2+m_i^2}}\frac{R(k)}{(k+p_i+p_j-p_i)^2-m_j^2},
\end{equation}
\noindent where $R(k)$ is the analytic rest of the integrand and
$k+p_i$ is taken on-shell. We now change the integration momentum to
the three-momentum of $k^{\mu\prime}=\Lambda^{\mu\nu}(k+p_i)_\nu$ in
the rest frame of $p_j-p_i$ and switch to spherical coordinates with
radial coordinate $\mathbf{k'}$
\begin{equation}\label{propinrestframe}
\int \frac{\mathbf{k'}^2 d\mathbf{k'}d\Omega}{2\sqrt{\mathbf{k'}^2+m_i^2}}\frac{R(\Lambda^{-1}k'-p_i)}
{m_i^2-m_j^2+2p_{ji}^0\sqrt{\mathbf{k'}^2+m_i^2}+{p_{ji}^0}^2}.
\end{equation}
\noindent Here we used the Lorentz invariance of the integration
measure $d^3k/2E_i$. In this system the peak lies on a surface of a
sphere with radius $\mathbf{k}_s$ given by (\ref{ks}). Expanding the
denominator around $\mathbf{k}_s$ yields
\begin{equation}
\int \frac{\mathbf{k'}^2 d\mathbf{k'}d\Omega}{2\sqrt{\mathbf{k'}^2+m_i^2}}\frac{R(\Lambda^{-1}k'-p_i)}
{\frac{2 p_{ji}^0 \mathbf{k}_s}{\sqrt{\mathbf{k'}_s^2+m_i^2}}(\mathbf{k'}-\mathbf{k}_s)+\mathcal{O}\left((\mathbf{k'}-\mathbf{k}_s)^2\right)}.
\end{equation}
Taking the limit $\mathbf{k}' \rightarrow \mathbf{k}_s$ the residue of the integrand is
\begin{equation}
\mbox{Res}(k'_s)=\frac{\mathbf{k_s}}{4p_{ji}^0}R(\Lambda^{-1}k'_s-p_i).
\end{equation}
Here, $k'_s$ is a four-vector with the spatial part fixed onto the surface with radius $\mathbf{k}_s$:
\begin{equation}\label{ks4}
k'_s=(\sqrt{\mathbf{k}_s^2+m_i^2},\mathbf{k}_s\frac{\vec{k'}}{|\vec{k'}|}).
\end{equation}
If we subtract
\begin{equation}
\frac{\mbox{Res}(k'_s)}{\mathbf{k'}-\mathbf{k}_s}
\end{equation}
from the integrand, this additional term does not add to the principal
value of the integral if it is integrated over a region with
symmetrical borders around the singular point $\mathbf{k}_s$. The peak
of the original integrand vanishes.

Thus, we can define a fixing function, which in the rest frame of $p_j-p_i$ reads
\begin{equation}\label{dff}
\mbox{Fix}(\mathbf{k'},k'_s)\equiv\frac{\mathbf{k}_s R(\Lambda^{-1}k'_s-p_i)}{4p_{ji}^0}\left(\frac{1}{\mathbf{k'}-\mathbf{k}_s}-2\frac{\mathbf{k'}-\mathbf{k}_s}{c^2}+
\frac{(\mathbf{k'}-\mathbf{k}_s)^3}{c^4}\right)
\theta(\mathbf{k'}-(\mathbf{k}_s-c))\theta((\mathbf{k}_s+c)-\mathbf{k'}),
\end{equation}
with $\theta(x)$ being the step function. Here, we also added a linear
and cubic term in $(\mathbf{k'}-\mathbf{k}_s)$, to make the joined
integrand continuous and differentiable at the artificial borders
introduced by the theta functions. Since the integration over the
radial coordinate $\mathbf{k'}$ runs from zero to infinity, the width
$c$ of this subtraction term can maximally be taken to be the radius
$\mathbf{k}_s$. In the numerical evaluation stable results were
obtained, when we took the width of support of the subtraction term
equal to the infrared cutoff, $c=\Delta E_s$.

Transforming back to the original momentum of integration, we get
\begin{equation}\label{ff}
\int d\mathbf{k}'d\Omega \mbox{Fix}(\mathbf{k'},k'_s)=\int \frac{d^3k'}{\mathbf{k'}^2}\mbox{Fix}(\mathbf{k'},k'_s)=
\int\frac{\Vert\Lambda\Vert d^3k}{{\overrightarrow{\Lambda(k+p)}}^2} \mbox{Fix}(|\overrightarrow{\Lambda(k+p)}|,k'_s(\overrightarrow{\Lambda(k+p)})),
\end{equation}
where $\overrightarrow{\Lambda(k+p)}$ is the spatial part of the
transformed four-vector and $\Vert\Lambda\Vert$ the corresponding
Jacobian.  In (\ref{ff}), we also indicated the dependence of the
fixed four-vector $k'_s$ on the spatial integration momentum
$\mathbf{k'}$ in the second argument of the fixing function.

When added to the integrand, the subtraction term (\ref{ff}) cancels
the peak which arises when a pole in the lower $k^0$ half plane and a
pole in the upper half plane fall together in the original loop
integrand. Note that the double delta term in (\ref{ftt}) including
$\Delta^l_i$ and $\Delta^u_j$ is supported at
$\mathbf{k'}=\mathbf{k}_s$ and adds an imaginary part to the result.

\subsubsection{Overlapping Peaks}
If propagator $P_i$ is cut, there might also exist a further
propagator $P_k$ fulfilling the conditions (\ref{condI}) and
(\ref{condII}) in addition to propagator $P_j$\footnote{The kinematical
  function $\lambda(p^2,m_1^2,m_2^2)$ of two adjacent loop propagators
  can only be positive if the momentum $p^2$ of the external particle
  is off-shell. For loop graphs with only on-shell external legs, the
  above possibility of two additional propagators getting singular at
  the same time is therefore given from six-point functions on. For
  $n_\textindex{off}$ off-shell external legs, this situation can only
  occur for loops with at least $6-n_\textindex{off}$
  propagators.}. If this is the case, another fixing function has to be
added to the same integrand smoothing the second peak. In a general
inertial frame the peaks have the form of rotational ellipsoids. In
principle, these two peaks may overlap, leading to a line in the
integration volume where both propagators $P_j$ and $P_k$ can get
singular at the same time. This is equivalent to a non-vanishing
contribution of a term including the three delta functions
$\Delta^l_i\Delta^u_j\Delta^u_k$ in (\ref{ftt}). In this case we have
to add a further fixing function.

The conditions for the occurrence of the two peaks, (\ref{condI}) and
(\ref{condII}), as well as the condition for an intersection of these
peaks can be checked numerically. When cutting propagator $P_i$ in the
rest frame of $p_j-p_i$, the radius of the sphere where propagator
$P_j$ gets singular is given by (\ref{ks}). With this we can construct
the on shell four-vector (\ref{ks4}). We now transform into the rest
frame of $p_k-p_i$:
\begin{equation}\label{ks4p}
{k^\mu_s}'=\Lambda^{\mu\nu}{k_s}_\nu \;,\;\;\;\;\; k_s^\mu=(\sqrt{\mathbf{k}_s^2+m_i^2},\mathbf{k}_s\frac{\vec{k}}{|\vec{k}|}).
\end{equation}
Similar to (\ref{propeq0}), propagator $P_k$ gets singular if
\begin{equation}
m_i^2-m_k^2+2\sqrt{p^2_{ki}}\frac{p^0_{ki}}{|p^0_{ki}|}\sqrt{\mathbf{k'}_s^2+m_i^2}+p_{ki}^2\cond 0,
\end{equation}
where $\mathbf{k'}_s$ is the absolute value of the spatial part of the
four-vector ${k^\mu_s}'$ in (\ref{ks4p}) and $p^0_{ki}$ is the
non-transformed zero component of $p_{ki}=p_k-p_i$. Using
\begin{equation}\label{Lk2}
{\overrightarrow{\Lambda(\beta)k}}^2=\gamma^2\left(k_0-\vec{\beta}\vec{k}\right)^2-m^2, \hspace{8mm}\mbox{with}\hspace{8mm}
\vec{\beta}=\frac{\vec{p}}{p^0}, \;\;\;\; \gamma=\frac{|p^0|}{\sqrt{p^2}},
\end{equation}
we get
\begin{equation}
m_i^2-m_k^2+2\sqrt{p^2_{ki}}\frac{p^0_{ki}}{|p^0_{ki}|}\gamma ({k}^0_s-\vec{\beta}\vec{k}_s)+p_{ki}^2\cond 0.
\end{equation}
Using $\gamma=\frac{|p^0_{ki}|}{\sqrt{p^2_{ki}}}$ and equation
(\ref{poscond1}) to replace ${k}^0_s$ we obtain:
\begin{equation}
\vec{\beta}\vec{k}_s\cond\frac{m_i^2-m_k^2+p_{ki}^2}{2p_{ki}^0}+\frac{m_j^2-m_i^2-p_{ji}^2}{2p_{ji}^0}.
\end{equation}
This condition is fulfilled if the righthand side is between the
bounds $+|\beta k_s|$ and $-|\beta k_s|$. If this is the case, we have
to add a third fixing function in the region where the two singularities
overlap. This will be shown in the following.

For better readability, we are now changing to a more symbolic
notation. Suppose the integrand has the form:
\begin{equation}\label{fabl1}
\frac{f(r,\theta,\phi)}{(r-a)(r'(r,\theta,\phi)-b)},
\end{equation}%
where $r,\theta,\phi$ are spherical coordinates in the integration
system and $r'$ a function of these coordinates which is the radial
coordinate in another coordinate frame. The function
$f(r,\theta,\phi)$ represents the non-singular rest of the
integrand. The fixing function which cancels the first peak can be
written as:
\begin{equation}\label{two_ff}
\mbox{Fix}_1=\frac{f(a,\theta,\phi)}{\overline{(r-a)}(r'(a,\theta,\phi)-b)}
\end{equation}
Here, the non-singular part is fixed at $r=a$. The line over the
factor in the denominator indicates that the fixing function is to be
subtracted from the original integrand in a region symmetric around
the corresponding singularity. One could imagine adding a second fixing
function to cancel the second peak similarly. In the region where the
two singularities overlap, one would naturally add a third function
which fixes the numerator to points on the intersection line and
projects each singular factor onto the singular surfaces of the other
factor:
\begin{equation}
\mbox{Fix}_3=\left.\frac{f(a,\theta,\phi)}{\overline{(r-a)}}\right|_{r'=b}\frac{1}{(\overline{r'(a,\theta,\phi)-b)}}.
\end{equation}
However, unless the two peaks are orthogonal to each other, this third
fixing function does not cancel the remaining peaks but gives even rise
to new singularities. These occur at points where the opening angle of
the two normals to the singular surfaces is small. Here, the two
factors in the denominator will become very small if projected onto
the surfaces. Being of second order in that small distance, this peaks
will not be cancelled by the fixing functions (\ref{two_ff}), where only
one of the two factors will be small at this point. Therefore, this
naive approach cannot be used.

We therefore have to find an alternative way to cancel the peaks. If
only the second singularity, $(r'-b)^{-1}$ in (\ref{fabl1}) was
present, the expression for the fixed integrand would read:
\begin{equation}\label{fix1}
\frac{f(r,\theta,\phi)-\left.f(r,\theta,\phi)\right|_{r'=b}}{(r'(r,\theta,\phi)-b)}.
\end{equation}
In the limit $r'\rightarrow b$, this is equivalent to the derivative
of $f$ with respect to $r'$ at $r'=b$ and we can interpret
(\ref{fix1}) as a result of an operation on $f$ similar to
differentiation without taking the limit $r'\rightarrow b$. If we
assume $f$ to be differentiable, no singularities are
introduced. Using the construction of the fixing function introduced in
the beginning of this section, expression (\ref{fix1}) is continuous
and differentiable. We can therefore again operate on (\ref{fix1}) to
get the difference equation with respect to $r$:
\begin{eqnarray}\label{fix123}
\frac{f(r,\theta,\phi)}{(r-a)(r'(r,\theta,\phi)-b)}-
\frac{\left.f(r,\theta,\phi)\right|_{r'=b}}{(r-a)(\overline{r'(r,\theta,\phi)-b)}}\nonumber\\
-\frac{f(a,\theta,\phi)}{\overline{(r-a)}(r'(a,\theta,\phi)-b)}+
\frac{\left.f(a,\theta,\phi)\right|_{r'=b}}{\overline{(r-a)}(\overline{r'(a,\theta,\phi)-b)}}.
\end{eqnarray}
This expression is again continuous and does not have any peaks. The
last three terms can therefore be interpreted as fixing functions to the
original integrand. However, the second and forth term are not zero
anymore. The factor $(r-a)$ is not fixed to a constant $r'$ and gives
asymmetric contributions if the fixing function is subtracted in a region
symmetric around $r'=b$. To get reliable results, the
region where these fixing functions are used should therefore be
small.  Our (arbitrary) choice of using the energy resolution $\Delta
E$ as the width of support of the fixing functions restricts the
corresponding error to the same magnitude as power corrections induced
by the standard treatment of IR cancellations.

The above analysis extends to any number of propagators which get
singular simultaneously. We can always start with a non-singular
function $f$ and apply the difference equation (\ref{fix1}), transform
into another frame and again using (\ref{fix1}) with respect to
another variable and so on. However, an accurate estimation of the
inflicted error by adding these fixing function is still missing.

\subsubsection{Higher-Order Fixing Functions}

The subtraction terms (\ref{t_op}) added by the renormalization scheme
to cancel the UV divergent terms sometimes contain squared
propagators. Performing the $k^0$ integration, one has to take the
derivative of the analytic rest of the integrand with respect to $k^0$
before replacing it according to the delta function obtained from the
cut propagator. It can happen that another propagator gets singular
and the resulting peak would be of second order. In this case one can
construct a further fixing function. The relevant term of the integrand
in the rest frame, similar to (\ref{propinrestframe}), is
\begin{equation}
I_2=\frac{\mathbf{k'}^2}{2\sqrt{\mathbf{k'}^2+m_i^2}}\frac{R(\Lambda^{-1}k'-p_i)}
{\left(m_i^2-m_j^2+2p_{ji}^0\sqrt{\mathbf{k'}^2+m_i^2}+{p_{ji}^0}^2\right)^2}.
\end{equation}
Expanding numerator and denominator separately around $\mathbf{k'}-\mathbf{k}_s$ we get
\begin{equation}
I_2=\frac{\sqrt{\mathbf{k}_s^2+m_i^2}}{8{p_{ji}^0}^2(\mathbf{k'}-\mathbf{k}_s)^2}
\frac{\mathbf{k}_sR(\mathbf{k}_s)+(2R(\mathbf{k}_s)+\mathbf{k}_sR'(\mathbf{k}_s))(\mathbf{k'}-\mathbf{k}_s)+\mathcal{O}((\mathbf{k'}-\mathbf{k}_s)^2)}
{\mathbf{k}_s+(\mathbf{k'}-\mathbf{k}_s)+\mathcal{O}((\mathbf{k'}-\mathbf{k}_s)^2)},
\end{equation}
where we wrote $R(\mathbf{k}_s)$ for
$R(\Lambda^{-1}k'_s-p_i)$. Multiplying by
$(\mathbf{k'}-\mathbf{k}_s)^2$ we get the coefficients of the poles in
the Laurent series by taking the limit $\mathbf{k'}\rightarrow
\mathbf{k}_s$ or taking the derivative with respect to $\mathbf{k'}$
and then taking the limit:
\begin{eqnarray}
I_2&=&\frac{r_{-2}}{(\mathbf{k'}-\mathbf{k}_s)^2}+\frac{r_{-1}}{(\mathbf{k'}-\mathbf{k}_s)}+\ldots,\\
&&\nonumber\\
r_{-2}&=&\frac{\sqrt{\mathbf{k}_s^2+m_i^2}}{8{p_{ji}^0}^2}R(\mathbf{k}_s),\\
r_{-1}&=&\frac{\sqrt{\mathbf{k}_s^2+m_i^2}}{8{p_{ji}^0}^2}\frac{R(\mathbf{k}_s)+\mathbf{k}_sR'(\mathbf{k}_s)}{\mathbf{k}_s}.
\end{eqnarray}
For the first order peak we can construct a fixing function equivalent to
(\ref{dff}) with the new residue $r_{-1}$. For the second order pole
we define the fixing function as:
\begin{equation}
\mbox{Fix}_2(\mathbf{k'},k'_s)\equiv r_{-2}\left(\frac{1}{(\mathbf{k'}-\mathbf{k}_s)^2}+\frac{2}{c^2}-\frac{3(\mathbf{k'}-\mathbf{k}_s)^2}{c^4}\right)
\Theta(\mathbf{k'}-(\mathbf{k}_s-c))\Theta((\mathbf{k}_s+c)-\mathbf{k'}).
\end{equation}
Here, the expression in the brackets results from taking the
derivative of the corresponding expression of an already fixed first
order function with respect to $\mathbf{k}$. Since we defined the
first order fixing function in (\ref{dff}) such that after addition to
the singular function the resulting expression is differentiable, also
the derivative does not develop a singularity. Transforming the two
fixing functions back to the original momentum frame and subtracting
them from the integrand removes the peaks. Note that applying the
difference equation (\ref{fix1}) twice at the same point in the same
coordinate frame would have lead to the same result.

\newpage
\section{Proof of Concept - Bhabha Scattering}
\label{s_proof_of_concept}
As a first application, we evaluate the one-loop cross section of
Bhabha scattering in massive QED. This $e^+e^-\rightarrow e^+ e^-$
scattering process is of great importance in electron-positron
colliders for the precise measurement of the luminosity at the
interaction point. For small angles, this process is dominated by the
kinematic singularity of the photon exchanged in the t-channel. The
differential cross section in this limit is proportional to the
scattering angle $\theta^{-4}$, and therefore gives a high event rate
and allows for a precise determination of the luminosity.  The
experimental accuracy aimed for the luminosity measurement at the
planned ILC is below $1 \permil$, cf. \cite{Brau:2007zza}. To match
this experimental precision, theoretical predictions of the Bhabha
scattering cross section should have at least the same
accuracy. Therefore, higher-order corrections have to be included in
the calculations and implemented in the Monte Carlo event
generators. The QED $\mathcal{O}(\alpha)$ corrections were calculated
long ago~\citerange{redhead,Berends:1973jb}, followed by the
one-loop electroweak corrections~\cite{Consoli:1979xw}.\footnote{To
  get an overview of the present status of Bhaba scattering
  calculations,
  cf.~\citerange{Montagna:1998sp,Actis:2007pn} and
  references therein.}  
%

%

The Bhabha scattering process is ideally suited to demonstrate the
evaluation of processes at NLO by the Feynman Tree Theorem.  The
one-loop result for the cross section is well known and can easily be
produced using automated loop-graph evaluation tools.  Real radiation
is well under control and can also be calculated and simulated using
automated tools.  On the FTT side, Bhabha scattering is a process
where most complications inherent in the method -- UV subtractions, IR
cancellations, and threshold singularities -- are present
simultaneously.  There are ten graphs in the one-loop corrections,
which after rewriting them as tree graphs lead to a rich structure in
the integrand.  This has to be treated by multichannel integration
methods.  Furthermore, the smallness of the electron mass compared to
the energy of $500\;\GeV$ where we evaluate the process provides a
stringent test of the stability of the numerical integration.

The results shown in the following include the virtual corrections as
well as the soft real emission parts. Addition of the hard emission part
would not change our results qualitatively and could be obtained straightforwardly 
in a multi purpose event generator. Nevertheless, the stated numerical
results for NLO cross sections should be interpreted bearing in mind the
missing hard real emission part.

\subsection{Implementation}

We created analytical expressions for the loop graphs in
computer-readable form using the Mathematica- and FORM-based
packages \feynarts\ and \formcalc\cite{Hahn:2000kx,Hahn:1998yk}. Here,
we blocked the reduction to tensor integrals, such that we obtained
squared matrix elements with the loop momentum still present in scalar
products. These matrix elements were then handed over to a specially
crafted Mathematica program that creates subtraction
graphs, cuts the loops and, where needed, calculates the fixing
functions. For each tree graph, the program creates parameterizations
(integration channels) which map the resulting phase space onto the
unit hypercube, taking into account the peak structure. The resulting
expressions for the matrix elements and the channels are then written
out in Fortran code.

As an integration routine, we choose the multi-channel algorithm
\vamp\cite{Ohl:1998jn}.  We compare the final results for the cross
section and angular distribution to an independent calculation that
proceeds along the usual way of integration via tensor reduction,
dimensional regularization, $\overline{\textrm{MS}}$ subtraction, and
numerical evaluation of the analytical result, using \feynarts\ and
\formcalc.

In case of the photon self-energies, we only consider the electron
loops. These corrections are infrared finite and have a transverse
Lorentz structure. Here, the final loop contribution is just a
correction factor to the Born matrix elements, and we use a compact
expression in terms of scalar integrals, which is then evaluated using
the Feynman Tree Theorem.
\begin{figure}[t]
\begin{center}
\begin{picture}(85,58)

\put(0,0){\begin{minipage}[b]{8.5cm}
\includegraphics[width=8.5cm]{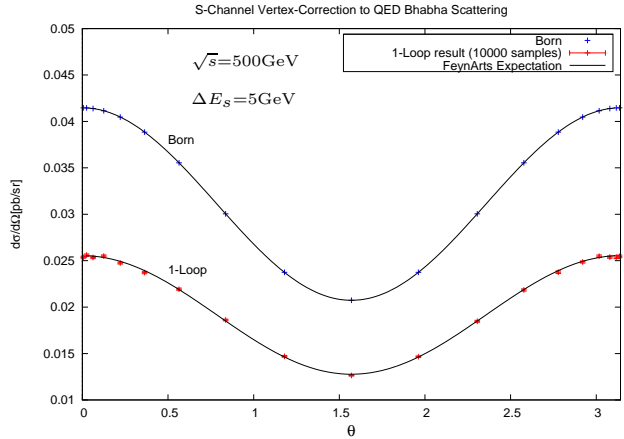}
\end{minipage}}
\put(25,47){$\begin{array}{l} {\scriptscriptstyle \sqrt{s}=500\mbox{{\tiny \GeV}} } \\
	{\scriptscriptstyle \Delta E_s=5\mbox{{\tiny \GeV}}} \end{array}$}
\end{picture}
\vspace{-3mm}
\caption{One-loop vertex correction to s-channel. Lines are obtained from \feynarts . Points are results from the Feynman Tree Theorem, numerically evaluated using \vamp .} \label{f_vc_s}
\end{center}
\end{figure}

\subsection{Integration Results}

In figure \ref{f_vc_s}, we show the differential cross section
resulting of a single vertex correction in the s-channel. For this and
the following plots of the differential cross section, we insert a
center of mass energy of $\sqrt{s}=500\GeV$ and an infrared soft
energy cutoff of $E_\textindex{soft}=5\GeV$. The results are in
complete agreement with \feynarts. This single graph already involves
a fixing function that cancels the threshold peak which corresponds to
the two leptons in the s-channel becoming on-shell. Furthermore, this
plot confirms the correct implementation of the ultraviolet
subtraction scheme proposed in sections \ref{s_ren_reg} and
\ref{s_uv/ir}.  Since, the leptons are back to back, there is
essentially only one collinear peak in the integrand. Nevertheless we
can deduce from the plot that the adaption of the grids in the
multi-channel approach to this peak works quite efficient.  With 10000
sampling points, the error estimate on the numerical integration as
returned by the Monte-Carlo integrator is less than $1\%$.

\begin{figure}[t]
\begin{center}
\begin{picture}(170,58)

\put(0,0){\begin{minipage}[b]{8.5cm}
\includegraphics[width=8.5cm]{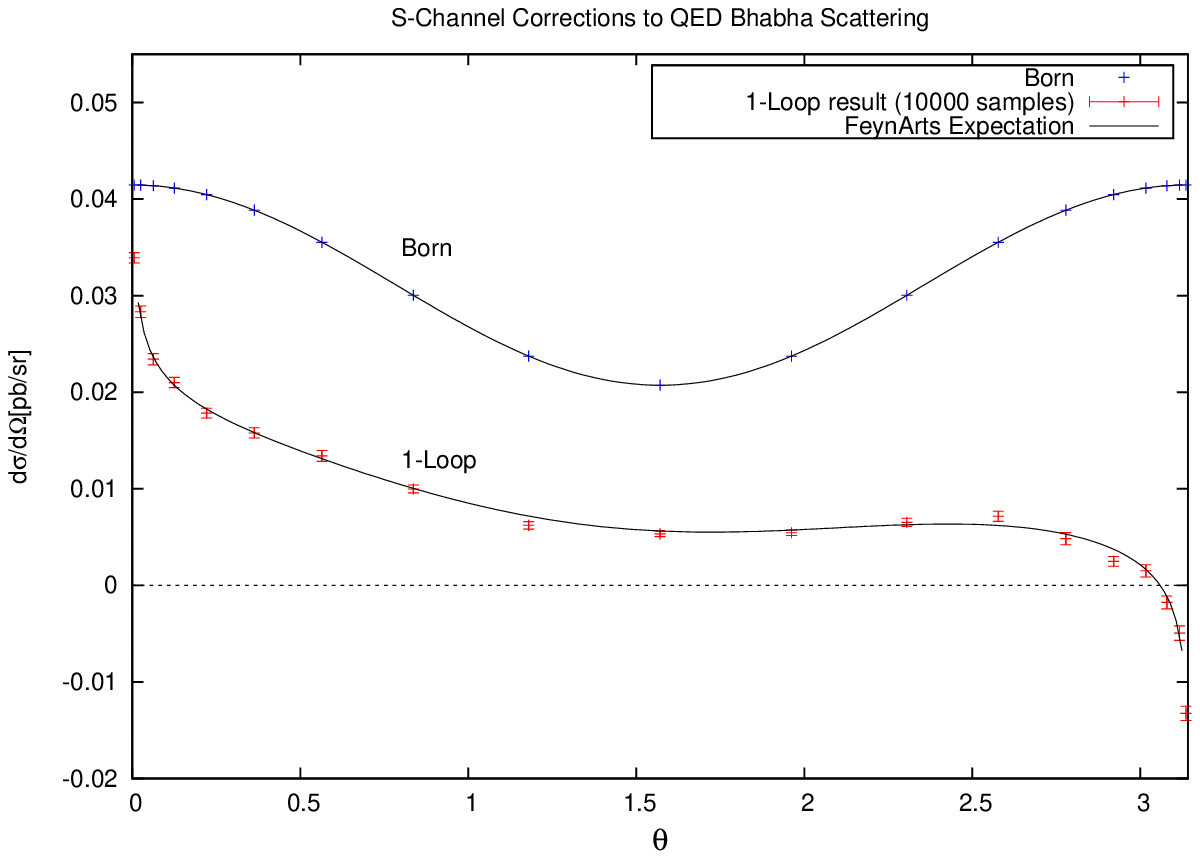}
\end{minipage}}\put(85,0)
{\begin{minipage}[b]{8.5cm}
\includegraphics[width=8.5cm]{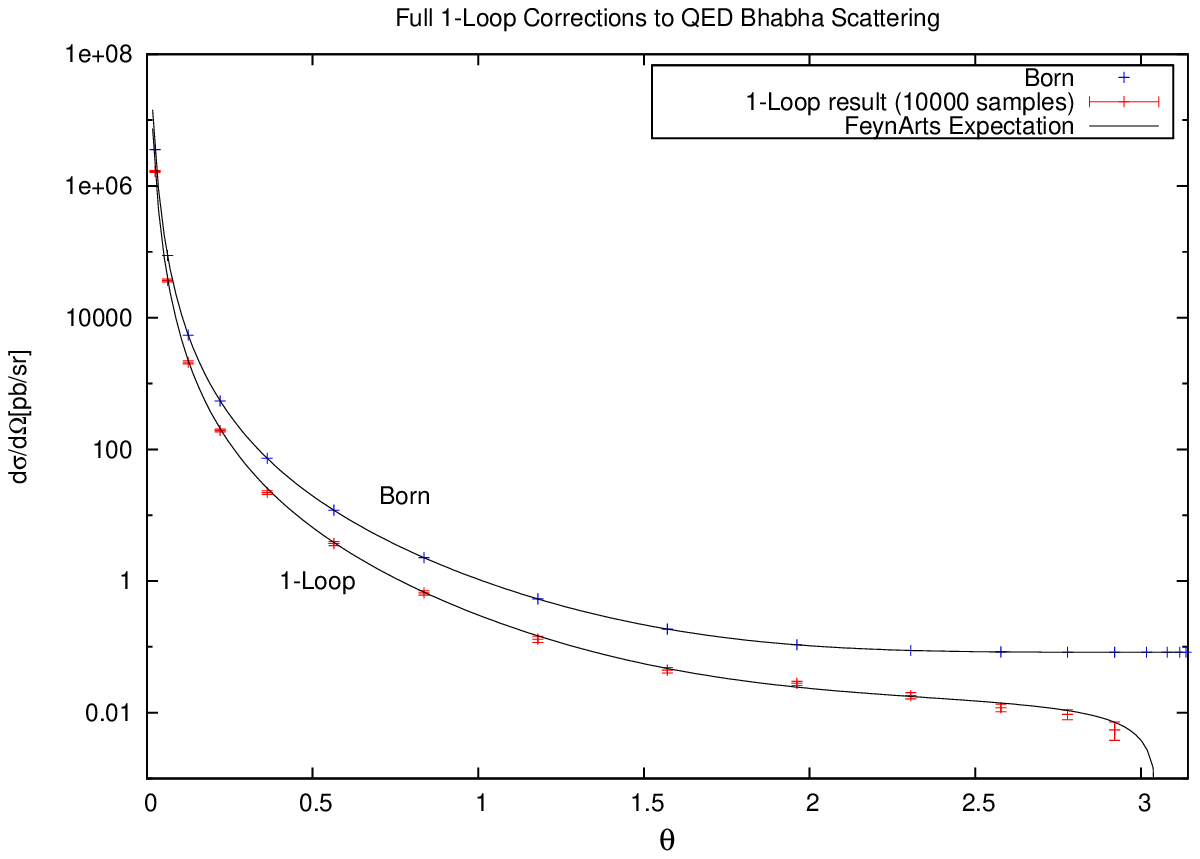}
\end{minipage}}

\put(10,10){$\begin{array}{l} {\scriptscriptstyle \sqrt{s}=500\mbox{{\tiny \GeV}} } \\
	{\scriptscriptstyle \Delta E_s=5\mbox{{\tiny \GeV}}} \end{array}$}
\put(105,47){$\begin{array}{l} {\scriptscriptstyle \sqrt{s}=500\mbox{{\tiny \GeV}} } \\
	{\scriptscriptstyle \Delta E_s=5\mbox{{\tiny \GeV}}} \end{array}$}

\end{picture}
\vspace{-3mm}
\caption{Differential cross section of $\mathcal{O}(\alpha)$-correction to s-channel Bhabha scattering (left) and the full $\mathcal{O}(\alpha)$-correction (right).
} \label{f_results_s/all}
\end{center}
\end{figure}
%
%
%
%
%
%

In figure \ref{f_results_s/all}, we separately show the differential
cross section for the complete s-channel contribution and the full
correction, since the s-channel is small compared to the t-channel
contribution in almost all of the phase space.  In both cases, there
are residual collinear peaks left in the integrand. As before, the
results from the multi-channel integrator have an error of
$\mathcal{O}(1\%)$.

\begin{figure}[t!]
\begin{center}
\begin{picture}(85,58)

\put(0,0){\begin{minipage}[b]{8.5cm}
\includegraphics[width=8.5cm]{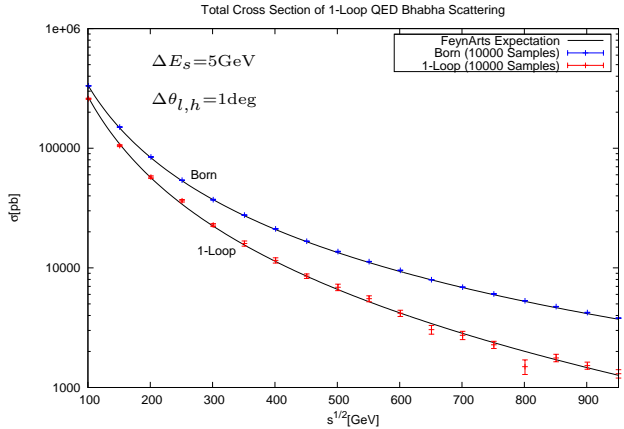}
\end{minipage}}
\put(20,47){$\begin{array}{l} 
	{\scriptscriptstyle \Delta E_s=5\mbox{{\tiny \GeV}}} \\
	{\scriptscriptstyle \Delta \theta_{l,h}=1\mbox{{\tiny deg}} }\end{array}$}

\end{picture}
\vspace{-3mm}
\caption{Integrated Cross Section of $\mathcal{O}(\alpha)$-corrections to QED Bhabha Scattering.} \label{f_results_ics}
\end{center}
\end{figure}

Figure \ref{f_results_ics} shows the total cross section as function
of the squared beam energy $s$. Results are again in complete agreement with
\feynarts.
%



\subsection{Event Generation}

So far, we have used multichannel integration for the purpose of
obtaining a numerically stable cross section and angular distribution.
As mentioned in section \ref{s_ftt}, we aim at using the Feynman Tree
Theorem to produce individual events at NLO level in a general purpose
Monte Carlo package. Here, we demonstrate the applicability by
constructing a partonic event generator for the Bhabha scattering
process.

The core of a Monte Carlo event generator is the hard partonic
sub-process. This is a $2\rightarrow n$ scattering matrix element with
$(3n-4)$ independent kinematic variables $x_i$. Any random choice of
these variables defines an event.  For the actual sampling, we make
use of the multi-channel integration routine \vamp, which distributes
the number of sampling points between different integration channels
based on their relative contributions to the variance, and
simultaneously adapts a discrete binning of each integration dimension
for each channel in order to internally minimize this variance
contribution.  We assign a weight $w_i$ to each sampled event, which
is the value of the integrand containing the original matrix element
multiplied by Jacobians arising form transformations between
integration channels and from grid adaption.  These weighted events
can be transformed into a sequence of unweighted events, i.e., physics
simulation, by a simple rejection algorithm.  A particular event is
accepted if a randomly chosen number $r \in [0,1]$ is lower than the
ratio of the weight $w_i$ of this event and a maximal weight
$w_\textindex{max}$.  The maximal event usually cannot be calculated
but has to be inferred from previous preparatory runs.

The resulting partonic events are subject to further refinements:
convolution of the initial state by structure functions, parton shower
and photon radiation, hadronization, and finally detector simulation.
These parts are not considered here.


Using the integrand obtained from the Feynman Tree Theorem, we will
generate unweighted events on the level of a hard partonic
sub-process, whose phase space includes the on-shell momentum of a
loop particle.  This loop particle is obviously unobservable, and its
momentum has to be integrated over when considering physical
observables.  However, this unphysical degrees of freedom opens the
possibility of negative event weights.  Each physical phase space
point corresponds to three extra dimensions of unphysical loop
momenta, and the events at this point will come both with positive and
negative weights.  In the average, the event weight at this point
is positive definite, but fluctuations are possible.

This source of negative event weights is distinct from the usual
problem of negative event weights near an IR/collinear singularity.
In that case, the problem can in principle be solved by a suitable
resummation prescription, although this is often not done or
technically impossible.  The physical event weight has to be positive
(precisely, unity), for each event, at \emph{any} phase-space point.
The negative event weights in our case, by contrast, are inherent in
the approach and cancel out in the average over most part of phase
space.  In the vicinity of IR singularities, our approach suffers from
the usual source of negative weights as well, since we have not
applied any resummation.  When unweighting our events, we make use of
the usual approach of unweighting events with positive and negative
weight separately, so we have an event sequence with weight of either
$+1$ or $-1$.

As explained above, in addition to the phase-space variables $x_i$ of
the $n$ final state particles we now simultaneously sample the three
variables $k_i$ of the original loop momentum. Thus, the number of
dimension of the integration is given by $(3n-1)$ for a $2\rightarrow
n$ process. Since the error inflicted by the acceptance-rejection
method is a purely statistical one, the addition of the phase space
variables of the inclusive (loop) particles does not influence the
error. However, the reweighting efficiency, which is given by the
number of kept events divided by the total amount sampled, will
generally decrease. In any new variable the grids of the integration
methods have to adapt to the shape of the integrand. It is therefore
helpful to use as much information about the peak structure as
possible to create the corresponding channels, which allow for an
efficient adaption of the integration grids.

We accept an event if
\begin{equation}
r\le\frac{|w_i|}{w^\pm_\textindex{max}},
\end{equation}
\noindent with random number $r \in [0,1]$ and
$w^\pm_\textindex{max}=\mbox{max}(|w_\textindex{max}|,|w_\textindex{min}|)$
being the absolute maximal weight encountered in the grid adaption and
integration steps. Each event was assigned an additional flag $\pm 1$
according to the sign of $w_i$. The inclusion of events with negative
weights increases the error as well as reduces the efficiency, since
the number of kept events is bigger than the effective number of
events, which is the difference of positive and negative events. 

\begin{figure}[t]
\begin{center}
\begin{picture}(170,58)

\put(0,0){\begin{minipage}[b]{8.5cm}
\includegraphics[width=8.5cm]{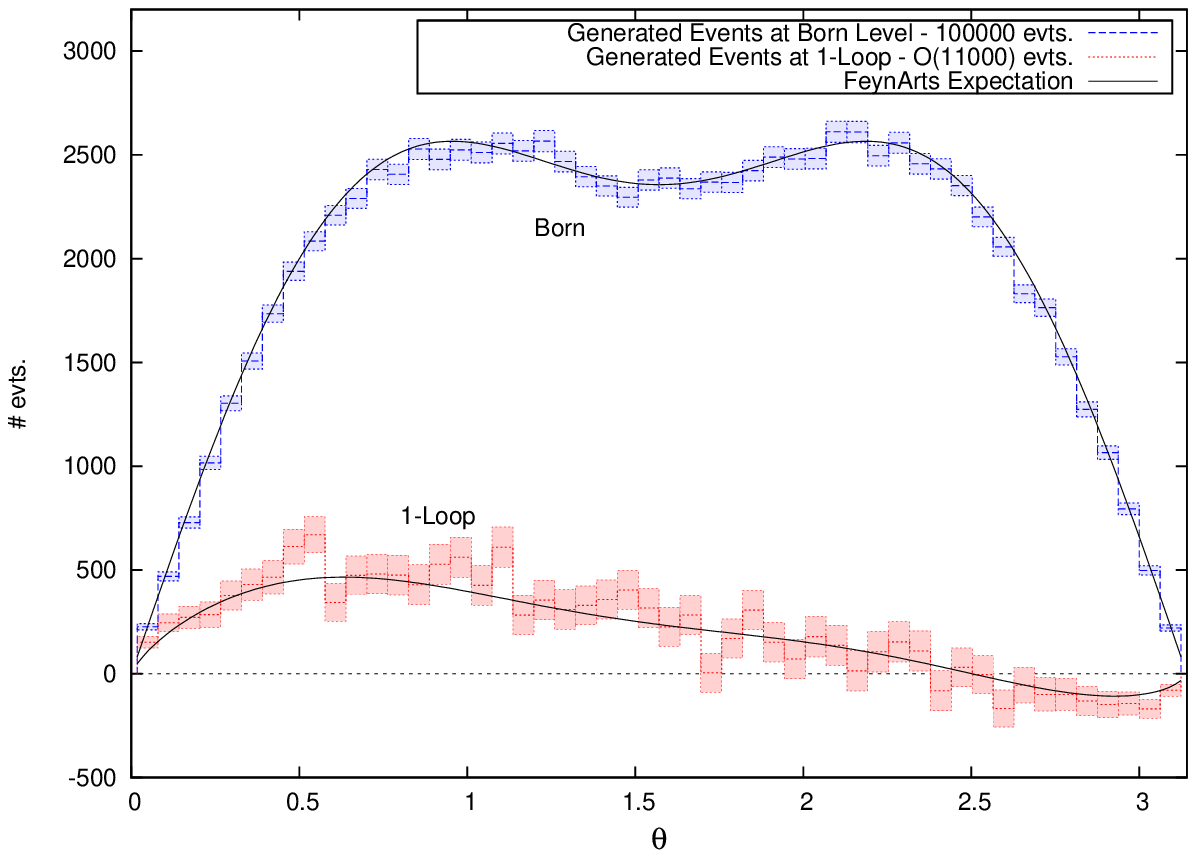}
\end{minipage}}\put(85,0)
{\begin{minipage}[b]{8.5cm}
\includegraphics[width=8.5cm]{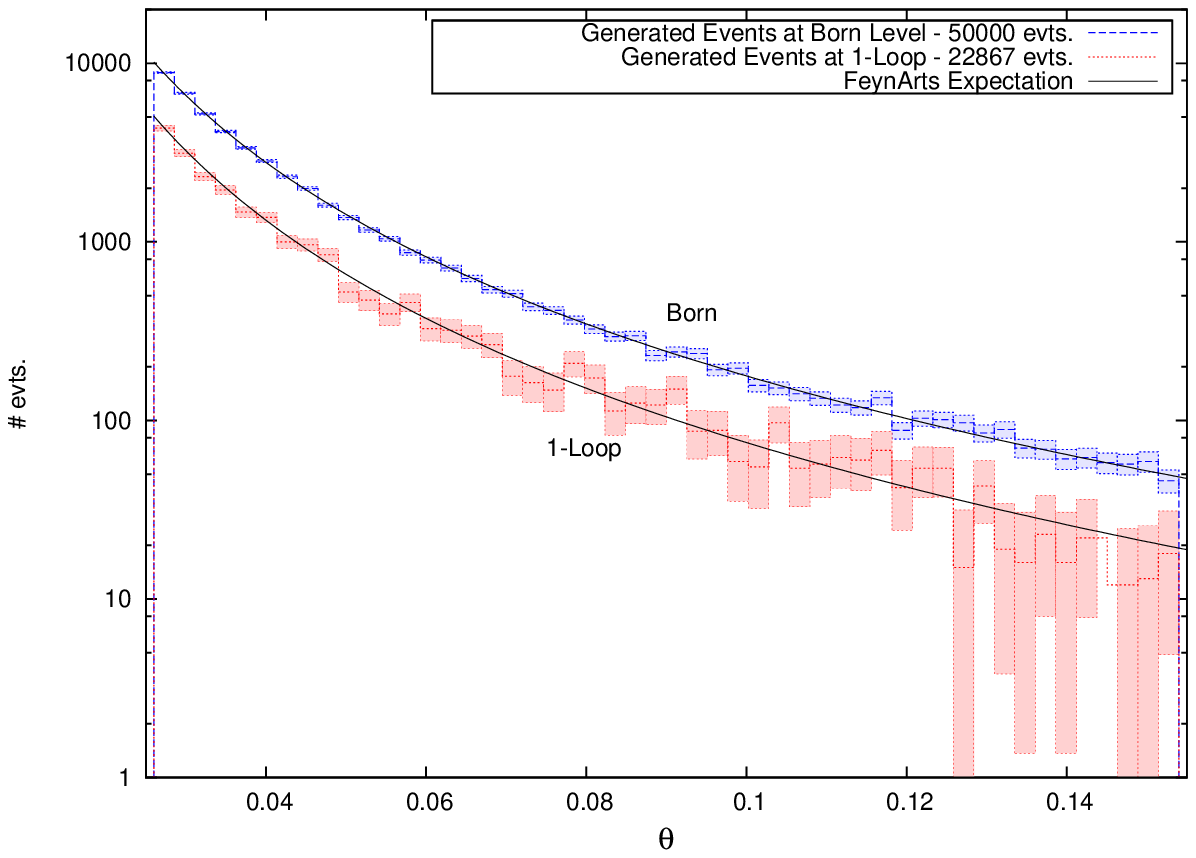}
\end{minipage}}
\put(10,52){$\begin{array}{l} {\scriptscriptstyle \sqrt{s}=500\mbox{{\tiny \GeV}} } \\
	{\scriptscriptstyle \Delta E_s=2.5\mbox{{\tiny \GeV}}} \end{array}$}
\put(98,13){$\begin{array}{l} {\scriptscriptstyle \sqrt{s}=500\mbox{{\tiny \GeV}} } \\
	{\scriptscriptstyle \Delta E_s=5\mbox{{\tiny \GeV}}} \end{array}$}
\end{picture}
\caption{Generated events for Bhabha scattering in the s-channel (left) and forwards scattering region (right). Born (blue/dashed) and NLO (red/dotted) results are shown with corresponding statistical error. Results are compared with the expectation from \feynarts\ (black/solid) calculation.} \label{f_evt_gen}
\end{center}
\end{figure}

In figure \ref{f_evt_gen}, we show results for Bhabha scattering
separately for the s-channel contribution, and for the complete result
in the forward scattering region.  We emphasize that these results
take only into account the LO and NLO partonic QED cross section
without any extras, while a generator of practical use should include
all effects from beamstrahlung and resummed ISR, resummed FSR, and NLO
electroweak contributions as well as NNLO QED contributions.
Nevertheless, for our numerical simulation we have adapted the
parameter settings taken for typical processes at a planned linear
collider \cite{Brau:2007zza,LDC}.

To demonstrate the handling of negative events, we used a small
infrared cutoff of $0.5\%$ of the center of mass energy
$\sqrt{s}=500\GeV$ in the s-channel contribution. This leads to a
differential cross section which is negative in parts of the phase
space. The total Born and NLO cross sections are obtained from Monte
Carlo integration of the grids set up for event generation:
\begin{equation}
\sigma^\textindex{tot}_\textindex{Born}=0.34745(29)\mbox{pb};\;\;\;\;\;
\sigma^\textindex{tot}_\textindex{NLO}=0.0338(58) \mbox{pb}.
\end{equation} 
Using only the Born level result, we generated 100000 unweighted events which corresponds to an integrated luminosity of about $\mathcal{L}=290 \mbox{fb}^{-1}$. Since the Born result is strictly positive, we did not encounter any negative weights. Here, the efficiency is
\begin{equation}
\mbox{eff}_\textindex{Born} = \frac{n^\textindex{evts.}}{n^\textindex{calls}}=61\%,
\end{equation}
where we adapted the grid in 6 iterations using 1000 samples each,
discarded the integral and performed an integration with a total
number of 15000 samples in three iterations. Comparing the NLO with
the Born cross section, we want to generate 9736 unweighted events. If
we allow for events with a negative weight, it follows that we have to
generate events until the difference of positive and negative events
equals 9736. Since in this case the integrand is rather equally
distributed among positive and negative values, we had to generate a
total amount of about 360000 events (186644 positive and 176908
negative). The efficiency for generating all events is
$\mbox{eff}^\textindex{p+n}_\textindex{NLO}=4.1\%$, however, the
efficiency of generating events which finally show up in the histogram
after the negative events are subtracted from the positive ones in
each bin, namely 11655 events, is at the per mil level:
$\mbox{eff}^\textindex{hist}_\textindex{NLO}=0.2\%$. Clearly, this is
a rather extreme case where a huge scale ratio ($\sqrt{s}$
vs.\ electron mass / energy resolution) enters the game, and the
differential cross section is small and even (without resummation)
negative in some regions of the phase space. Therefore, it is natural
that the integrand is spread among positive and negative values which
hampers an efficient event generation. 

Here, further manipulation of the integrand like dipole subtraction,
or mapping the negative onto the positive parts might improve the
behavior.  Nevertheless, even without such additional improvements,
the results are reliable and stable.

The right-hand plot of figure \ref{f_evt_gen} shows events in the
forward scattering region. The covered region corresponds to
$\theta_\textindex{min}=26\mbox{mrad}$ and
$\theta_\textindex{max}=154\mbox{mrad}$ \cite{LDC}. The integrated
cross sections for the Born and the NLO process are:
\begin{equation}
\sigma^\textindex{tot}_\textindex{Born}=5981.9(2.3) \mbox{pb};\;\;\;\;\;\sigma^\textindex{tot}_\textindex{NLO}=2736(82) \mbox{pb}.
\end{equation}
We generated 50000 events for the Born process with an efficiency of
$\mbox{eff}_\textindex{Born} =72\%$, requiring about 23000 events for
the NLO process. These were accepted with an efficiency of
$\mbox{eff}^\textindex{p+n}_\textindex{NLO}=11\%$, dropping down to
$\mbox{eff}^\textindex{p-n}_\textindex{NLO}=1.8\%$, accounting for the
effect of negative events.

\section{Conclusions}

We have developed a new method for computing NLO corrections to
scattering cross sections.  The method is based on the Feynman tree
theorem and bears some similarities to other methods which draw on
cuts and analycity properties of Feynman integrals.  However, our
approach differs in the fact that all integrals are transformed into
ordinary phase-space integrals (albeit with unusual boundaries) that
can be handled by an ordinary numerical multi-channel phase-space
integrator.  To this hand, we had not just to implement subtractions
for UV and IR singularities, but furthermore subtraction functions
(fixing functions) for threshold singularities which do not cause
problems in the usual semi-analytic methods.

By computing the complete result for a well-known process and
constructing an unweighted NLO event generator, we could show that the
method actually works, even in kinematically difficult regions of
parameter space.  The resulting code could be added to a standard
tree-level event generator, and it is straightforward in principle to
extend the method to other processes, or even handle them
automatically.

Extending the method to the full Standard Model and multi-particle
processes, it promises three important advantages over more
conventional semi-analytic algorithms: (i) While the evaluation of a
simple process such as Bhabha scattering is clearly more complicated
than by standard methods, the computational complexity does not
increase dramatically with the number of legs in loop diagrams.
Evaluating a NLO $n$-particle process should require similar CPU
resources as a LO $n+1$-particle process, summed inclusively over all
particle species.  This is handled regularly by standard universal
event generators.  (ii) The presence of masses in loop graphs does not
worsen the performance.  Instead, it improves the numerical stability.
Therefore, we expect the method to be most useful for models with many
mass scales (such as the MSSM), once a proper definition of
subtractions and renormalization conditions has been set up.  (iii)
Combining loop integration and phase-space sampling in a single step,
we avoid a whole layer in the calculation.  In particular, all terms
are evaluated only up to the level of precision that is required by
the actual simulation.

However, there is still a long way before this method can actually
improve the simulation of physics processes in the Standard Model or
its extensions.  On the one hand, we have to handle the more
complicated IR behavior of QCD and state suitable renormalization
conditions for the non-abelian theory.  On the other hand, the method
has to be augmented by a consistent treatment of unstable states such
as $W$ and $Z$ bosons, which appear in loops and, in our case, would
become artificial external particles in event samples.  Nevertheless,
the method, if it can be applied to the complete Standard Model has
distinct advantages that warrant its further development towards
realistic complete LHC and ILC applications.

          
\subsection*{Acknowledgements}

This work has been supported in part by the
Helmholtz-Hochschul-Nachwuchsgruppe VH--NG--005 and 
the UK Science and Technology Facilities Council (STFC), PP/E 0073717/1.\\
TK would like to thank Z. Kunszt and T. Kennedy for useful discussions.

          
 


\baselineskip15pt

\end{document}